\documentclass[lettersize,journal]{IEEEtran}
\usepackage{amsmath,amsfonts}
\usepackage{amssymb}
\usepackage{algorithmic}
\usepackage{algorithm}
\usepackage{array}
\usepackage{subcaption}
\usepackage{amsthm}
\usepackage{textcomp}
\usepackage{stfloats}
\usepackage{url}
\usepackage{xcolor}
\usepackage{verbatim}
\usepackage{graphicx}
\usepackage{flushend}
\usepackage{cite}
\usepackage{hyperref}
\usepackage{booktabs}
\allowdisplaybreaks
\begin{document}

\title{
3D Gaussian Radiation Field Modeling for Integrated RIS-FAS Systems: Analysis and Optimization
}

\author{ 
Kaining Wang,~\IEEEmembership{Student Member,~IEEE,} 
Bo~Yang,~\IEEEmembership{Senior Member,~IEEE,} Yusheng Lei, Zhiwen Yu,~\IEEEmembership{Senior Member,~IEEE,}  Xuelin Cao,~\IEEEmembership{Senior Member,~IEEE,} Liang Wang,~\IEEEmembership{Member,~IEEE,} Bin Guo,~\IEEEmembership{Senior Member,~IEEE,} George C. Alexandropoulos, \IEEEmembership{Senior Member,~IEEE,}  M\'erouane Debbah,~\IEEEmembership{Fellow,~IEEE},  
and Zhu Han,~\IEEEmembership{Fellow,~IEEE}
 \thanks{K. Wang, Y. Lei, B. Yang, L. Wang, and B. Guo are with the School of Computer Science, Northwestern Polytechnical University, Xi'an, Shaanxi, 710129, China (email: wangkaining@mail.nwpu.edu.cn, yang$\_$bo, liangwang, guob@nwpu.edu.cn). 

Z. Yu is with the School of Computer Science, Northwestern Polytechnical University, Xi'an, Shaanxi, 710129, China, and Harbin Engineering University, Harbin, Heilongjiang, 150001, China (email: zhiwenyu@nwpu.edu.cn).

 X. Cao is with the School of Cyber Engineering, Xidian University, Xi'an, Shaanxi, 710071, China (email: caoxuelin@xidian.edu.cn). 


 G. C. Alexandropoulos is with the Department of Informatics and Telecommunications, National and Kapodistrian University of Athens, 16122 Athens, Greece (email: alexandg@di.uoa.gr). 


M. Debbah is with  KU 6G Research Center, Department of Computer and Information Engineering, Khalifa University, Abu Dhabi 127788, UAE (email: merouane.debbah@ku.ac.ae).

Z. Han is with the Department of Electrical and Computer Engineering at the University of Houston, Houston, TX 77004 USA, and also with the Department of Computer Science and Engineering, Kyung Hee University, Seoul, South Korea, 446-701(email: hanzhu22@gmail.com).

}
}





\maketitle

\begin{abstract}
The integration of reconfigurable intelligent surfaces (RIS) and fluid antenna systems (FAS) has attracted considerable attention due to its tremendous potential in enhancing wireless communication performance. However, under fast-fading channel conditions, rapidly and effectively performing joint optimization of the antenna positions in an FAS system and the RIS phase configuration remains a critical challenge. Traditional optimization methods typically rely on complex iterative computations, thus making it challenging to obtain optimal solutions in real time within dynamic channel environments. To address this issue, this paper introduces a field information-driven optimization method based on three-dimensional Gaussian radiation-field modeling for real-time optimization of integrated FAS–RIS systems. In the proposed approach, obstacles are treated as virtual transmitters and, by separately learning the amplitude and phase variations, the model can quickly generate high-precision channel information based on the transmitter’s position. This design eliminates the need for extensive pilot overhead and cumbersome computations. On this framework, an alternating optimization scheme is presented to jointly optimize the FAS position and the RIS phase configuration. Simulation results demonstrate that the proposed method significantly outperforms existing approaches in terms of spectrum prediction accuracy, convergence speed, and minimum achievable rate, validating its effectiveness and practicality in fast-fading scenarios.
\end{abstract}


\section{Introduction}

\IEEEPARstart{W}{ith} the rapid evolution of large-scale intelligent wireless networks, the sixth-generation (6G) mobile communications are envisioned to realize the ultimate goal of intelligent connectivity and programmable radio environments. Unlike conventional systems that passively adapt to wireless propagation, 6G networks emphasize cognition and reconfigurability, enabling communication nodes not only to perceive but also to proactively reshape their surroundings for dynamically optimal transmission.
Among various enabling technologies, the reconfigurable intelligent surface (RIS) has emerged as a promising paradigm to achieve environment-controllable communications~\cite{0}. By reprogramming the phase shifts of its passive reflecting elements, RIS can redirect incident electromagnetic (EM) waves toward desired directions without requiring additional radio-frequency (RF) chains, thereby reconstructing the propagation environment in a cost- and energy-efficient manner~\cite{1,3,4}. 

However, the performance of RIS-assisted systems in dynamic scenarios is fundamentally limited by channel non-stationarity and pilot overhead. Owing to the passive and large-scale nature of RIS, spectral leakage, phase coupling, and reflection misalignment in multi-user or multi-path environments can significantly degrade system throughput~\cite{5,6}. Existing RIS optimization schemes rely heavily on iterative channel estimation and matrix-based optimization, whose computational complexity grows exponentially with the number of reflecting elements, thus making real-time configuration infeasible under fast time-varying channels~\cite{7,8,8-1}.

\subsection{Motivation} 
To overcome the limitations of passive reflection, the fluid antenna system (FAS) has been proposed as a flexible antenna architecture capable of physically moving its antenna within a small spatial region~\cite{9,11-1,11-2}. By continuously selecting the most favorable reception position, FAS exploits small-scale spatial diversity to enhance the desired signal and suppress interference, all without additional bandwidth or power consumption. Motivated by the complementary strengths of RIS and FAS, their integration establishes a new communication paradigm that enables joint adaptability across transmission, propagation, and reception. In this framework, the RIS controls large-scale wavefront shaping, whereas the FAS refines the received signal through spatial sampling and position optimization~\cite{10,10-1,10-2,11}. This hierarchical coordination enables dynamic trade-offs between spectral efficiency, coverage uniformity, and interference suppression.

Nevertheless, the FAS–RIS paradigm still faces several challenges. In particular, the positional mobility of the FAS introduces a continuous spatial dimension to the channel state information (CSI), making accurate characterization of the radiation field across space increasingly demanding. Consequently, conventional pilot-based estimation may lead to excessive training overhead, especially when fine spatial granularity is required. Moreover, the joint optimization of RIS phase shifts and FAS positions involves multi-scale nonconvex coupling between reflection control and spatial field distribution, which further complicates system design. While analytical and ray-tracing models provide valuable insights into the propagation characteristics, they are often limited in capturing the fine spatial correlations inherent in FAS-assisted systems. Meanwhile, emerging neural implicit representations, such as NeRF \cite{12} or volumetric field models, offer high expressive power but usually at the expense of computational efficiency and physical interpretability ~\cite{13}.

To address these limitations, this paper proposes a data-driven optimization framework based on the three-dimensional (3D) Gaussian radiation field (3DGRF) model. Unlike implicit neural fields that rely on dense sampling and heavy training, the 3DGRF represents the electromagnetic energy distribution using explicit Gaussian primitives. Each primitive encodes local amplitude and phase variations in a differentiable, physically consistent form, thereby enabling the efficient reconstruction of the continuous radiation field from limited observations. This explicit, interpretable representation enables analytical differentiation for gradient-based optimization of both the RIS phase configuration and FAS position control. Consequently, the proposed framework transforms traditional CSI-dependent adaptation into radiation-field-driven sensing and control, achieving low-complexity, real-time, and cognition-driven optimization for next-generation FAS–RIS systems.

\subsection{Related Work}
\subsubsection{Optimization of RIS–FAS Systems}
RIS can programmatically manipulate wireless propagation via passive phase control, while FAS enables dynamic antenna-position switching within a confined region to exploit spatial diversity and suppress interference. Their integration introduces a new class of environment-adaptive architectures for 6G networks. 

Early studies primarily focused on theoretical modeling of RIS–FAS systems. For example, Lai \textit{et al.}~\cite{14} developed an analytical framework for outage probability under a block-correlation model, while Yao \textit{et al.}~\cite{15} introduced a block-diagonal matrix approximation to simplify FAS port correlation and derive outage bounds. After this, Yao \textit{et al.}~\cite{16} proposed joint optimization under both full and partial CSI conditions, showing that CSI-free methods can outperform conventional approaches in high-mobility environments. These works provide valuable theoretical foundations but rely on simplified or statistical channel assumptions, limiting their ability to capture spatial non-stationarity and continuous field variation.

From the optimization perspective, Tang \textit{et al.}~\cite{17} minimized transmit power in multi-user RIS–FAS systems via alternating optimization of beamforming, RIS phase shifts, and FAS positions, assuming ideal CSI and static geometry. Yao \textit{et al.}~\cite{18} compared the performance of FAS and adaptive RIS (ARIS), showing their complementarity in different channel conditions. In terms of signal design, Zhu \textit{et al.}~\cite{19} proposed an FAS-based index modulation scheme for RIS-assisted mmWave systems, while Ghadi \textit{et al.}~\cite{20} analyzed FAS-based RSMA and STAR-RIS systems under phase mismatch, revealing that FAS can mitigate phase and hardware impairments. Similarly, \cite{21} demonstrated the potential of RIS–FAS integration for physical-layer security and blockage resilience by leveraging additional spatial degrees of freedom. Despite these advances, most studies remain confined to small-scale or idealized channel models and lack scalable, data-driven methods for spatially continuous field inference.

\subsubsection{Wireless Channel Modeling}
Accurate wireless channel modeling lies at the core of modern communication system design. Probabilistic models capture large-scale path loss and small-scale fading but neglect spatial correlation and directional propagation characteristics. Deterministic methods, such as ray tracing (RT) \cite{21-1}, offer physically accurate multipath modeling but incur exponential computational costs as scene complexity increases, limiting their applicability in dynamic environments. With the emergence of machine learning and computer vision, data-driven modeling has gained traction. NeRF-based methods~\cite{12} and NeWRF~\cite{13} reconstruct 3D radiation fields from sparse samples, while physically informed models such as WinERT~\cite{22}, WiseGRT~\cite{23}, and FIRE~\cite{24} combine physical priors with neural representations to improve generalization. Other approaches, such as R2F2~\cite{25} and OptML~\cite{26}, employ cross-frequency CSI prediction and differentiable channel estimation to enhance scalability. 

Nevertheless, these models exhibit an inherent trade-off between accuracy, interpretability, and computational efficiency. The ray tracing and NeRF-based frameworks achieve high fidelity but lack real-time capability, whereas lightweight neural or statistical models compromise on spatial realism. To bridge this gap, this paper introduces a 3DGRF framework tailored for RIS–FAS systems. By representing EM energy distributions with Gaussian primitives, the proposed approach unifies RIS reflection modulation and FAS spatial sampling into a continuous, differentiable radiation field. This physically grounded formulation transforms channel modeling from matrix-based estimation to field-driven optimization, achieving real-time reconstruction, interpretability, and adaptability in complex 6G wireless environments.

\subsection{Contributions and Organization}

We detail our contributions as follows:
\begin{itemize}
    \item A novel channel modeling approach based on 3D Gaussian (3DGS) power-spectrum reconstruction is proposed and applied to maximize the minimum achievable rate. Unlike traditional methods that rely on extensive pilot-based estimation, the proposed framework learns the 3D Gaussian power-spectrum distribution of the environment to achieve compact, efficient channel reconstruction. This approach enables the rapid generation of high-fidelity channel information with limited training overhead, providing a reliable foundation for maximizing the minimum achievable rate. In this way, the dependence on pilot signals is significantly reduced, while the system robustness and adaptability under fast-fading conditions are greatly enhanced.
    \item A comprehensive analysis of pilot overhead and interference management in the FAS–RIS architecture is conducted for the first time. While RIS enhances signal coverage, it inevitably introduces spectral leakage and interference coupling. FAS, on the other hand, mitigates such interference by dynamically adjusting the receiver antenna position. This work quantitatively investigates the pilot requirements and interference suppression capability of the FAS–RIS system under various propagation conditions, revealing the underlying synergy that enables joint performance improvement. The findings provide new insights into the feasibility and potential of deploying FAS–RIS in future large-scale communication systems.
    \item An efficient joint optimization algorithm is designed to achieve collaborative configuration of RIS and FAS using power-spectrum priors. The proposed algorithm fully exploits the prior information derived from the reconstructed 3DGS power spectrum, thereby avoiding the computational burden of high-dimensional iterative searches in conventional optimization. By incorporating power-spectrum priors as constraints in the optimization process, the proposed method achieves near-optimal phase control and position selection with low computational complexity, ensuring real-time adaptability and scalability under fast-fading conditions.
    \item Extensive simulations validate the effectiveness and superiority of the proposed framework. The results demonstrate that the proposed 3DGS reconstruction model significantly outperforms existing methods in channel reconstruction accuracy while achieving higher spectral efficiency, lower computational complexity, and reduced pilot overhead. Moreover, joint optimization of FAS and RIS effectively mitigates interference and enhances system capacity and fairness in multi-user scenarios.
\end{itemize}

 The rest of this paper is organized as follows. Section {\color{blue}\ref{s2}} presents the proposed system model and problem formulation; Section {\color{blue}\ref{s3}} introduces the 3DGRF Gaussian radiation-field modeling; Section {\color{blue}\ref{s4}} describes the Field-Driven Alternating Optimization (FAO) algorithm for joint FAS-RIS configuration; Section {\color{blue}\ref{s5}} reports the simulation results and performance comparisons. Finally, Section {\color{blue}\ref{s6}} concludes the paper.

\textit{Notations:} Scalars are denoted by uppercase italics unless specified otherwise; vectors and matrices are denoted by bold italic lowercase and bold uppercase letters, respectively. The sets of complex and real numbers are denoted by $\mathbb{C}$ and $\mathbb{R}$, respectively. The following notations are used throughout this paper: $\mathbb{C}$ denotes the complex field; $(\cdot)^H$ denotes the conjugate transpose; $|\cdot|_2$ denotes the $\ell_2$-norm.
For matrix $\boldsymbol{X}$, $\boldsymbol{X}^\top$ and $\boldsymbol{X}^{\sf H}$ denote transpose and Hermitian. $\Re\{\cdot\}$, $\Im\{\cdot\}$ extract real/imaginary parts. For vector $\boldsymbol{x}$, $\|\boldsymbol{x}\|_2$ is the $\ell_2$-norm and $\|\boldsymbol{X}\|_{\mathrm F}$ is Frobenius norm.

\begin{figure}
    \centering
    \includegraphics[width=\linewidth]{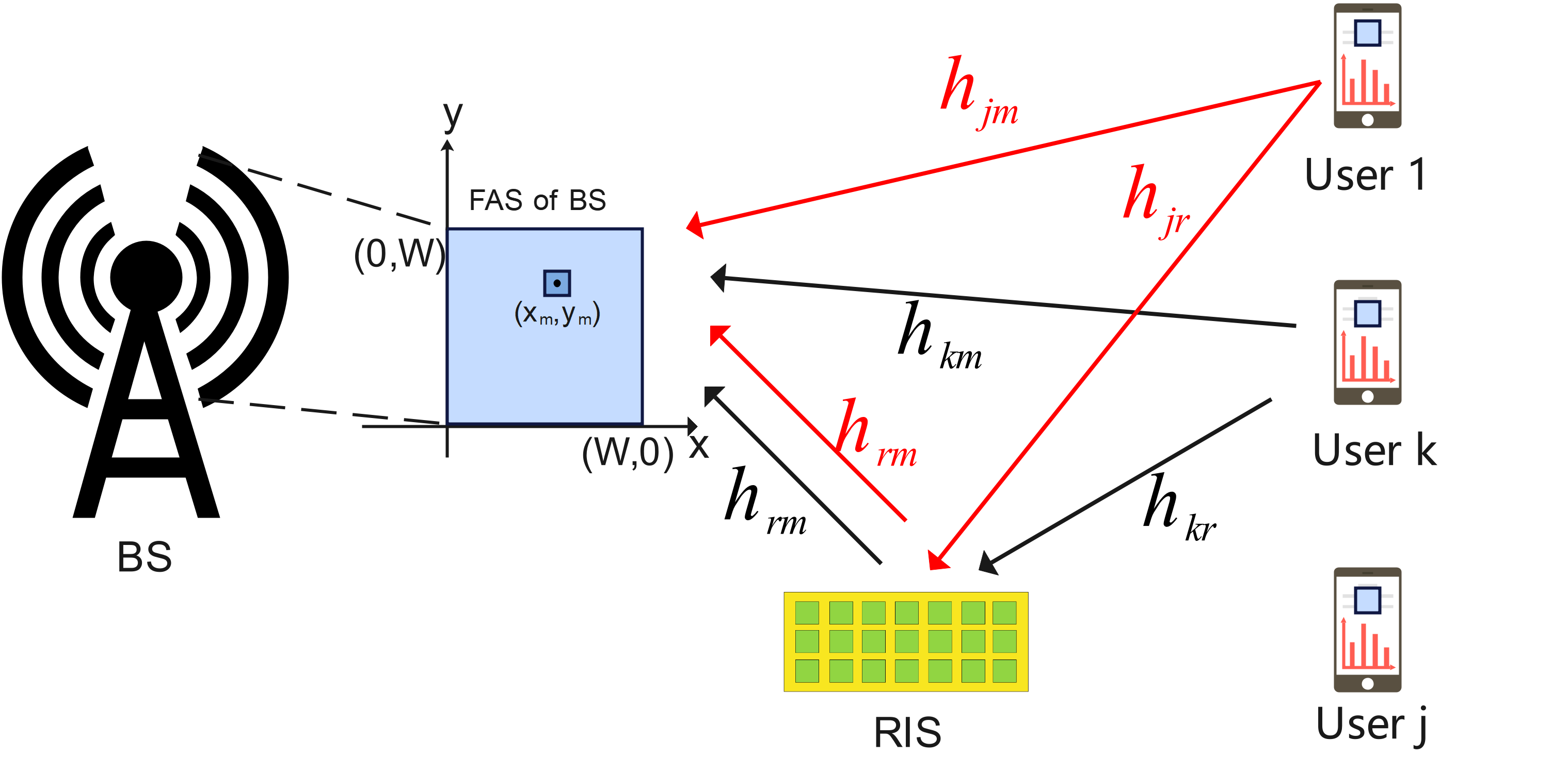}
    \caption{The considered system model of integrated RIS–FAS–assisted 
uplink multi-user communications. }
    \label{fig:scene}
\end{figure}

\section{ System Model and Problem Formulation}\label{s2}

\subsection{Signal Modeling}
As illustrated in Fig.~\ref{fig:scene}, we consider a multiuser FAS-RIS-assisted uplink communication system, where a total of $U$ users transmit signals simultaneously toward the base station (BS). Among them, user $k$ denotes the desired user, while other users $j$ act as interference sources, where $k \neq j$. 

Each user is equipped with a fixed-position antenna (FPA), whereas the BS has $M$ two-dimensional FASs. We define $\overline{\xi} = [\xi_1, \xi_2, \ldots, \xi_M]$ to represent the antenna positions at the BS side. Within each FAS, every antenna element can instantaneously switch to any location within the region $S_\xi = [0, W] \times [0, W]$.

The received signal at the $m$-th BS antenna from user $k$ can be expressed as
\begin{equation}
y_m = \sqrt{P} \, \mathbf{h}_{rm} \boldsymbol{\Theta} \mathbf{h}_{kr}^H + \sigma^2,
\label{eq:received_signal}
\end{equation}
where $P$ denotes the transmit power and $\sigma^2$ represents the additive white Gaussian noise (AWGN) power at the BS receiver.

The RIS consists of $N$ reflecting elements, capable of flexibly controlling both the amplitude and phase of the reflected signals to manipulate the propagation direction of electromagnetic waves. The cascaded reflection coefficient matrix is denoted by
\begin{equation}
\boldsymbol{\Theta} = \mathrm{diag}(\phi_1,\phi_2,\ldots,\phi_N),
\label{eq:E_matrix}
\end{equation}
where $\phi_n=exp(j\theta_n)$, and $n \in \mathcal{N} = \{1, \ldots, N\}$, where $\theta_n \in [0,2\pi]$.

\subsection{Channel Modeling}

Since the RIS is deployed close to the user side, the RIS–UE link is dominated by line-of-sight (LoS) propagation. However, the channels between the fluid antennas (FAs) and the surrounding nodes, namely the UE–FA and RIS–FA links, exhibit spatially correlated characteristics rather than independent random fading. This is because the FA can move within a small local region, where the wireless channel varies smoothly with spatial position. Modeling these channels as spatially correlated functions of the FA position ensures that optimizing the FA location yields physically consistent and meaningful performance gains.

Let the $m$-th FA be located at position $\mathbf{\xi}_m = [x_m, y_m]^T$. The RIS–FA channel between the RIS and FA $m$ is modeled as
\begin{equation}
\mathbf{h}_{rm} = \sqrt{\alpha_{\mathrm{rm}}\mathbf{R}_{\mathrm{rm}}} \, \mathbf{g}_{\mathrm{rm}},
\label{eq:hrm_spatial}
\end{equation}
where $\alpha_{\mathrm{rm}}$ denotes the large-scale path loss, $\mathbf{R}_{\mathrm{rm}}$ represents the spatial correlation matrix \cite{matrix,matrix2} describing the correlation of the RIS–FA link, and $\mathbf{g}_{\mathrm{rm}}$ is a small-scale fading vector whose phase varies smoothly with the FA position.

The spatial correlation matrix $\mathbf{R}_{\mathrm{rm}}$ can be modeled as
\begin{equation}
[\mathbf{R}_{\mathrm{rm}}] = \rho_{\mathrm{rm}}^{|\xi_m-\xi_m'|}, \quad 0 \leq \rho_{\mathrm{rm}} \leq 1,
\label{eq:RRF_model}
\end{equation}
where $\rho_{\mathrm{rm}}$ characterizes the spatial correlation between adjacent RIS–FA subchannels. A larger $\rho_{\mathrm{rm}}$ indicates stronger correlation due to the limited movement range of the FA within the local area.

Similarly, the UE–FA link is expressed as
\begin{equation}
\mathbf{h}_{km} = \sqrt{\alpha_{\mathrm{km}}\mathbf{R}_{\mathrm{km}}} \mathbf{g}_{km},
\label{eq:hkm_spatial}
\end{equation}
where $\alpha_{\mathrm{km}}$ is the path loss of the UE–FA link, $\mathbf{R}_{\mathrm{km}}$ is the corresponding spatial correlation matrix, and $\mathbf{g}_{\mathrm{km}}$ denotes the normalized local scattering vector dependent on FA position.

In the above expression, $\sqrt{\mathbf{R}_{km}}$ is the Hermitian square root, which captures the spatial dependency among the UE–FA subchannels. It can be modeled using an exponential correlation structure as
\begin{equation}
[\mathbf{R}_{\mathrm{km}}] = \rho_{\mathrm{km}}^{|\xi_m-\xi_m'|}, \quad 0 \leq \rho_{\mathrm{km}} \leq 1,
\label{eq:RUF_model}
\end{equation}
where $\rho_{\mathrm{km}}$ denotes the spatial correlation coefficient that reflects the similarity of fading between neighboring FA positions. The term $\sqrt{\mathbf{R}_{\mathrm{km}}}$ ensures that the generated channel vectors maintain the desired correlation pattern while preserving the overall power normalization.

Consequently, the composite effective channel observed by the UE is given by
\begin{equation}
\mathbf{h}_k
= \mathbf{h}_{km} + \mathbf{h}_{kr}\boldsymbol{\Theta}\mathbf{h}_{rm},
\label{eq:heff_spatial}
\end{equation}
where $\mathbf{h}_{kr}$ denotes the LoS RIS–UE channel.

\subsection{Problem Definition}
Based on the above signal and channel models, the signal-to-interference-plus-noise ratio (SINR) at BS antenna $m$ can be expressed as
\begin{equation}
\Gamma =
\frac{ \left| \sqrt{P} \, \mathbf{h}_{rm} \boldsymbol{\Theta} \mathbf{h}_{kr}^H \right|^2 }
{ \sum\limits_{j=1, j \neq k}^{U}
\left| \sqrt{P} \, \mathbf{h}_{rm} \boldsymbol{\Theta} \mathbf{h}_{jr}^H \right|^2 + \sigma^2 }.
\label{eq:sinr}
\end{equation}
Accordingly, the minimum achievable rate from multiple users to the BS is defined as
\begin{equation}
R = \log_2(1 + \Gamma).
\label{eq:rate}
\end{equation}

Therefore, the objective is to maximize the minimum achievable rate by jointly optimizing the RIS reflection coefficient matrix $\boldsymbol{\Theta}$ and the BS antenna positions $\overline{\xi}$. The optimization problem can be formulated as
\begin{equation}
\label{eq:opt_problem}
\begin{aligned}
\max_{\overline{\xi}, \boldsymbol{\Theta}} \quad & R  \\
\text{s.t.} \quad 
& \overline{\xi} \in S_\xi, \\
& \| \xi_m - \xi_v \|_2 \ge D, \quad m, v \in \mathcal{M}, \, m \neq v, \\
& \boldsymbol{\Theta} = \mathrm{diag}\{\phi_1, \ldots, \phi_N\}, \\
& \phi_n = \exp(j \theta_n), \quad \theta_n \in [0, 2\pi),
\end{aligned}
\end{equation}
where $D$ denotes the minimum inter-antenna spacing to avoid mutual coupling. However, due to strong coupling among the optimization variables in both the objective function and the constraints, as well as the problem's highly nonconvex nature, solving it poses significant challenges.

\section{3DGS-BASED RADIATION FIELD RECONSTRUCTION} \label{s3}
\begin{figure*}
    \centering
    \includegraphics[width=\linewidth]{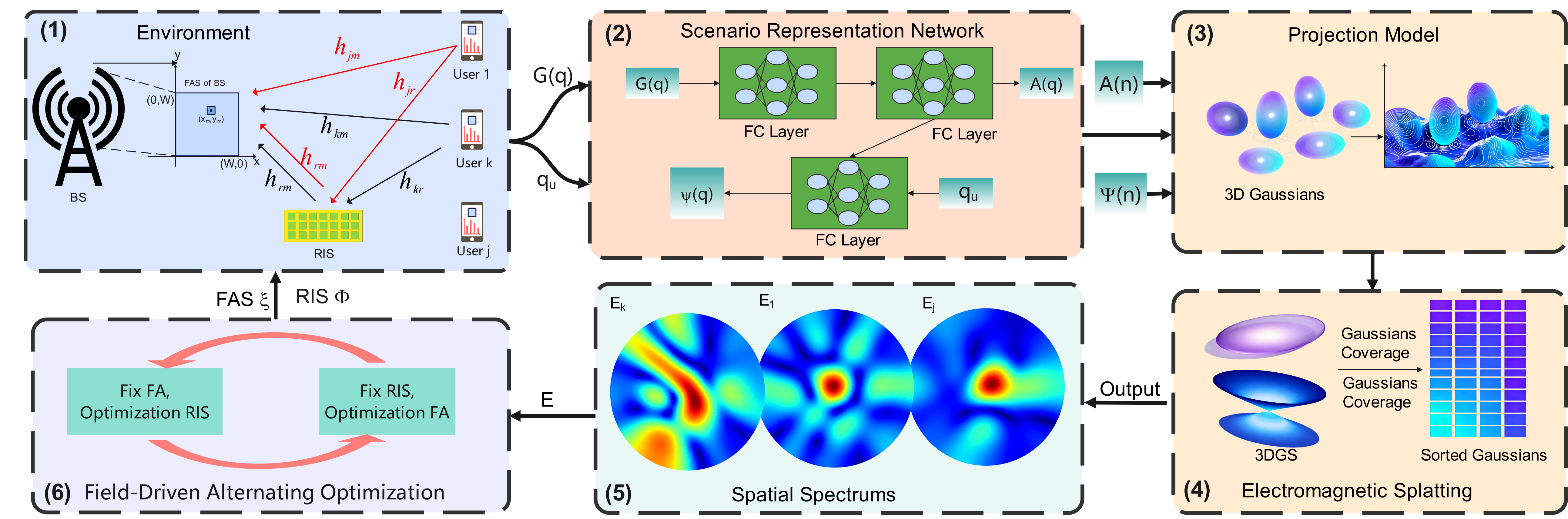}
    \caption{Overall framework of the proposed 3DGRF–based optimization. 
(1) The environment setup includes BS, RIS, and FAS. 
(2) The scenario representation network encodes geometric and transmitter information into latent radiation-field features. 
(3) The projection model represents the electromagnetic field using differentiable 3DGS primitives. 
(4) The electromagnetic splatting process reconstructs the continuous radiation field and generates spatial power distributions. 
(5) The reconstructed field produces spatial spectrums for subsequent system optimization. 
(6) The FAO iteratively updates FAS positions and RIS phases based on 3DGRF, achieving field-based control.}
    \label{fig:Reconstruction}
\end{figure*}

\subsection{Preliminaries}
Accurate modeling and reconstruction of wireless channels constitute the foundation for realizing environment-adaptive communication systems. However, conventional channel modeling approaches encounter significant limitations in scenarios involving the integration of RIS and FAS. The main challenge arises from the fact that RIS–FAS systems simultaneously involve multi-scale spatial wavefront manipulation and receiver-side position adaptation. Consequently, the channel characteristics dynamically vary with spatial locations and reflection states, making it difficult for static statistical or deterministic models to provide effective characterization.

Traditional probabilistic models rely on empirical formulas and statistical fitting, which can only capture average properties such as path loss and large-scale fading but fail to represent the fine-grained electromagnetic field distribution in space. As a result, these models exhibit considerable errors in multiuser and multipath RIS scenarios. Deterministic models, such as ray tracing, offer high physical interpretability but suffer from exponential computational complexity as environmental complexity increases. Moreover, they heavily depend on precise geometric modeling and material parameters, thus making real-time application in dynamic scenarios infeasible. In addition, deep-learning-based implicit channel modeling methods (e.g., the NeRF family) can learn nonlinear mappings between environmental geometry and signal propagation. Nevertheless, their training and rendering overheads are prohibitively high, preventing their practical use for real-time channel estimation and configuration in FAS–RIS systems.

In this context, the introduction of 3DGS provides an efficient and physically consistent paradigm for channel modeling in RIS–FAS systems. Originating from radiance field modeling in computer vision, 3DGS can reconstruct high-fidelity spatial field distributions from a limited number of sampled points in continuous 3D space. Compared with implicit representations such as NeRF, 3DGS adopts explicit Gaussian primitives, achieving orders-of-magnitude advantages in both training and rendering efficiency. More importantly, it facilitates the integration of electromagnetic propagation physics, enabling joint reconstruction of amplitude and phase information.


\subsection{Continuous Radiation Field Description}
Traditional channel models typically represent the wireless channel as the superposition of a finite number of discrete propagation paths, expressed as
\begin{equation}
h(\mathbf{\xi}) = \sum_{\ell=1}^{L} A_\ell e^{j \zeta_\ell},
\label{eq:discrete_paths}
\end{equation}
where $L$ denotes the number of propagation paths, $A_\ell$ and $\mathbf{\zeta}_\ell$ represent the complex gain and wave vector of the $\ell$-th path, respectively.
The implicit assumption is that $L$ is limited and that the received signal variations arise only from the interference among these discrete components.  

However, in practical scenarios—especially in environments involving RIS and spatially correlated FAs—the number of scattering and reflection paths can be extremely large, and their spatial distributions are often continuous rather than discrete.  
This observation aligns with the spatially correlated channel model discussed in the previous subsection, in which the received signal varies smoothly with the FA position.  

To capture this physical continuity, the wireless channel can be described as a continuous radiation field rather than a finite-path summation, i.e.,
\begin{equation}
h(\xi) = \int_{\Omega} A(\ell) e^{j 2\pi \zeta_\xi} \, d\mathbf{\ell},
\label{eq:continuous_field}
\end{equation}
where $A(\mathbf{\xi})$ denotes the spatial field spectrum associated with the spatial frequency component $\mathbf{\ell}$, and $\Omega$ represents the angular domain.  
This formulation bridges the discrete multipath representation and the spatially correlated field model, providing a unified and physically consistent framework for describing channel variations in FAS–RIS systems.  

This continuous field representation further lays the foundation for the subsequent field-driven optimization of RIS and FAS configurations.

To characterize the continuous distribution of energy and phase evolution in 3D space, we define a complex-valued radiation field function $E(\mathbf{\xi})$, representing the complex field strength at an arbitrary spatial observation position:

\begin{equation}
E(\mathbf{\xi}) = \sum_{\ell=1}^{L} A_\ell e^{j\theta_\ell} e^{-j\frac{2\pi}{\lambda}\|\mathbf{\xi}-\mathbf{q}_\ell\|},
\label{eq:virtual_emitter_model}
\end{equation}
where $\mathbf{\xi}$ denotes the spatial observation position (FA location), $\mathbf{q}_l$ denotes the position of the virtual emitter corresponding to the $l$-th path, and $A_l$ and $\theta_l$ represent its amplitude attenuation and phase shifts, respectively. This model no longer explicitly traces individual paths but instead describes the electromagnetic field distribution at any location through a continuous complex field $E(\mathbf{\xi})$. Such a representation transforms the propagation environment from a “set of discrete paths” to a “spatially continuous field,” a key step toward unified field-based modeling.

\subsection{3D Gaussian Primitives}
To approximate the continuous radiation field with a finite set of parameters, we employ 3DGS Primitives to model the field in a distributed manner. Originally developed for scene rendering in computer graphics, Gaussian primitives efficiently represent continuous spatial distributions of light energy with a small number of parameters. When adapted to the wireless communication domain, they can effectively describe both the amplitude and phase distributions of electromagnetic energy in 3D space. The 3DGS approach directly approximates the radiation energy cloud using analytical Gaussian functions, thereby avoiding volumetric integration and significantly reducing rendering and training overhead. This explicit representation naturally supports differentiable optimization, making it highly suitable for real-time updates of dynamic parameters in FAS-RIS systems.

The $i$-th Gaussian primitive, located at position $\boldsymbol{q}_i$, is defined as:
\begin{equation}
G_i(\mathbf{\xi}) = A_i e^{j\zeta_i}
\exp\!\left(-\frac{1}{2} (\mathbf{\xi} - \boldsymbol{q}_i)^{\mathrm{T}}
\boldsymbol{\Sigma}_i^{-1} (\mathbf{\xi} - \boldsymbol{q}_i) \right),
\label{eq:gaussian_primitive}
\end{equation}
where $A_i$ denotes the amplitude coefficient of the Gaussian primitive, reflecting its reflection or radiation strength, and $\zeta_i$ represents the associated phase factor. The covariance matrix $\boldsymbol{\Sigma}_i$ determines the spatial spread and anisotropy of the primitive\cite{wrfgs}.

The overall radiation field is obtained by superimposing all $N_G$ Gaussian primitives, i.e.,
\begin{equation}
\tilde{E}(\mathbf{\xi}) = \sum_{i=1}^{N_G} G_i(\mathbf{\xi}),
\label{eq:gaussian_field}
\end{equation}
where $N_G$ denotes the total number of Gaussian primitives used to approximate the field distribution. A larger $N_G$ allows for a finer representation of complex radiation patterns, while a smaller $N_G$ yields a more compact but coarser approximation.  

This formulation mathematically corresponds to a kernel expansion of the spatial energy density function and, physically, can be interpreted as the coherent superposition of electromagnetic fields generated by multiple virtual emission points.

\subsection{Gaussian Mapping Modeling for RIS-FAS Systems}
Since the reflection positions and phase shifts of each unit are fixed after configuration, they can be directly mapped to deterministic Gaussian primitive parameters. For the $n$-th RIS element, the corresponding phase and spatial center parameters are defined as:
\begin{equation}
\psi_n = \theta_n + \frac{2\pi}{\lambda} \left( d_{k,n} + d_{\xi,n} \right), \quad
\boldsymbol{\mu}_n = \mathbf{q}_n,
\label{eq:ris_gaussian_params}
\end{equation}
where $d_{k,n}$ and $d_{\xi,n}$ denote the propagation distances from the transmitter and receiver to the $n$-th RIS element, respectively, $\mathbf{r}_n$ represents the 3D spatial coordinates of the element, $\psi_n$ is the phase parameter of the $n$-th Gaussian primitive (determined by the RIS phase shift $\theta_n$ and propagation distances), and $\boldsymbol{\mu}_n$ is the spatial center of the primitive.

Accordingly, the RIS-reflected field can be expressed as a coherent superposition of Gaussian primitives:
\begin{equation}
E_{r}(\mathbf{\xi}) = \sum_{n=1}^{N}
A_n e^{j\psi_n}
\exp\!\left( -\frac{1}{2} (\mathbf{\xi}-\boldsymbol{\mu}_n)^{\mathrm{T}}
\boldsymbol{\Sigma}_n^{-1} (\mathbf{\xi}-\boldsymbol{\mu}_n) \right),
\label{eq:ris_field}
\end{equation}
where $A_n$ is the amplitude coefficient of the $n$-th RIS element (reflecting its reflection strength), and $\boldsymbol{\Sigma}_n$ is the covariance matrix determining the spatial spread of the primitive.

This formulation explicitly incorporates both the phase parameter $\psi_n$ and spatial position $\boldsymbol{\mu}_n$ of each RIS element into the Gaussian mapping process, thereby bridging the physical RIS configuration and the analytical 3D Gaussian field representation. When the RIS configuration is updated, only the phase shift $\theta_n$ (which determines $\psi_n$) and amplitude $A_n$ need to be adjusted to rapidly obtain the new field distribution without re-solving the entire channel.

The FAS receiver antennas can move within a constrained spatial region $S_\xi$. The received signal of the FAS is denoted as:

\begin{equation}
y_m = \sqrt{P} \, \tilde{E}(\mathbf{\xi}_m) + \sigma^2,
\quad \xi_m \in S_\xi,
\label{eq:fas_sampling}
\end{equation}
where $\mathbf{\xi}_m$ denotes the instantaneous spatial position of the $m$-th FAS antenna within the allowed region.

\subsection{Scenario Representation Network}
Geometric parameters alone are insufficient to characterize the complex environmental effects on electromagnetic propagation. Therefore, we introduce a Scenario Representation Network (SRN) to learn the signal-propagation behavior, following the deepSDF structure \cite{SRN}.

The SRN takes as input the transmitter position $\mathbf{p}_{\mathrm{tx}}$ and an environmental point-cloud coordinate $\mathbf{q}$, and outputs a pair of complex attenuation parameters $(\mu(\mathbf{q}),\delta(\mathbf{q}))$, formulated as:
\begin{equation}
\mathcal{F} : (\mathbf{q}, \mathbf{p}_{\mathrm{tx}})
\rightarrow (\mu(\mathbf{q}),\delta(\mathbf{q})).
\label{eq:srn_mapping}
\end{equation}

For each Gaussian point, the complex coefficient is expressed as:
\begin{equation}
C(\mathbf{q}) = \mu(\mathbf{q}) e^{j\theta(\mathbf{q})}.
\label{eq:srn_complex_coeff}
\end{equation}

The SRN consists of two multilayer perceptrons (MLPs). The first MLP extracts spatial geometric features and learns the attenuation distribution associated with the environmental position, while the second MLP fuses these features with the transmitter position to predict the signal's amplitude and phase responses.

This structure ensures the adaptability of Gaussian parameters to environmental geometry, enabling rapid generation of new Gaussian parameters under varying RIS reflection configurations or obstacle layouts.

The network is trained by minimizing the structural-similarity-aware reconstruction loss between the reconstructed and the ground-truth fields. Let $E_{\mathrm{gt}}(\mathbf{\xi})$ denote the measured ground-truth radiation field at spatial observation position $\mathbf{\xi}$, and $\tilde{E}(\mathbf{\xi})$ represent the field reconstructed by the SRN. We have the loss function as 
\begin{equation}
\mathcal{L} \!=\! (1 \!-\! \eta)\|E_{\mathrm{gt}}(\mathbf{\xi}) \!-\! \tilde{E}(\mathbf{\xi})\|_2^2
\!+\! \eta \big(1 \!-\! \mathrm{SSIM}(E_{\mathrm{gt}}(\mathbf{\xi}), \tilde{E}(\mathbf{\xi}))\big),
\label{eq:srn_loss}
\end{equation}
where $\eta$ is a weighting factor, and the SSIM is the Structural Similarity Index Measure function.
\begin{proof}
    See Appendix A.
\end{proof}

After training, the SRN can generate the corresponding Gaussian radiation field for any given RIS configuration and FAS position, enabling fast, physically consistent channel reconstruction.

\subsection{Projection Model and Electromagnetic Splatting}
The projection model maps the virtual TXs $\mathbf{q}$, represented by 3D Gaussians, onto the FA's perception plane, $\xi$.
To obtain the angular-domain response of the FAS antenna array, the 3DGRF must be projected onto the array’s perceptual plane. Since the receiving coverage of the array corresponds to a hemispherical region, we employ a Mercator projection to map spatial coordinates $(x_q, y_q, z_q)$ into angular coordinates $(\Omega_{\mathrm{lon}}, \Omega_{\mathrm{lat}})$:
\begin{equation}
\begin{cases}
\Omega_{\mathrm{lon}} = \arctan2(y_q, x_q), \\
\Omega_{\mathrm{lat}} = \arcsin\!\left(\dfrac{z_1}{\sqrt{x_1^2 + y_1^2 + z_1^2}}\right),
\end{cases}
\label{eq:mercator_projection}
\end{equation}
where $\Omega_{\mathrm{lon}}$ and $\Omega_{\mathrm{lat}}$ represent the longitude and latitude of the angular domain, respectively.

The angular space is then discretized to obtain the spatial power spectrum matrix:
\begin{equation}
P(\Omega_{\mathrm{lon}}, \Omega_{\mathrm{lat}})
= \big|\tilde{E}(\mathbf{\xi}(\Omega_{\mathrm{lon}}, \Omega_{\mathrm{lat}}))\big|^2.
\label{eq:angular_power_spectrum}
\end{equation}

Furthermore, at the implementation level, an efficient accumulation is achieved through the Electromagnetic Splatting mechanism. Physically, electromagnetic splatting characterizes the energy attenuation and phase accumulation of multipath signals at different propagation depths, serving as the radio-frequency counterpart of optical "light blending".

After each Gaussian primitive is projected onto the 2D angular plane, its contribution is accumulated sequentially according to depth order. Let $C_q(\Omega_q)$ denote the complex signal of the $q$-th Gaussian primitive at angular position $\Omega_q$. The total received signal is given by:
\begin{equation}
R_k = \sum_{i=1}^{N_G}
\left( \prod_{j=1}^{i-1} 
\mu(\mathbf{q}_j) e^{j\delta(\mathbf{q}_j)} \right)
C_q(\Omega_q),
\label{eq:em_splatting_signal}
\end{equation}
and the corresponding power spectrum is:
\begin{equation}
I(\Omega_k) = |R_k|^2.
\label{eq:em_splatting_power}
\end{equation}

This parallel splatting process is implemented on GPUs, allowing millisecond-level generation of spatial power spectra and enabling real-time visualization of the FAS–RIS channel.

By integrating the Gaussian mapping of RIS reflection elements and the spatial sampling of the FAS receiver, the complete system radiation field can be formulated as:
\begin{equation}
E_{k}(\mathbf{\xi}) =
\sum_{i=1}^{J} A_i e^{j\theta_i}
\exp\!\left(-\tfrac{1}{2}(\mathbf{\xi}-\mathbf{q}_i)^{\mathrm{T}}
\boldsymbol{\Sigma}_n^{-1}(\mathbf{\xi}-\mathbf{q}_i)\right),
\label{eq:fas_ris_field}
\end{equation}
where $J$ is the number of 3D Gaussians near a given pixel.
\begin{proof}
    See Appendix B.
\end{proof}

The received signal at the $m$-th FAS antenna position $\mathbf{\xi}_m \in S_\xi$ is expressed as:
\begin{equation}
y_m = \sqrt{P}\, E_{k}(\mathbf{\xi}_m) + \sigma^2.
\label{eq:fas_ris_received}
\end{equation}

\section{Proposed FAS-RIS Design}\label{s4}
In this section, to effectively tackle the joint optimization problem formulated in (\ref{eq:opt_problem}), we decompose it into two interdependent sub-problems: the continuous optimization of the FAS positions $\xi$ and the discrete optimization of the RIS reflection phase matrix $\Theta$. Leveraging the differentiable 3DGRF model, we propose a FAO framework that iteratively updates these two sets of variables to achieve a locally optimal solution.

\subsection{Problem Reformulation}
Building upon the 3DGRF formulation established in Section~\ref{s3}, we now reformulate the joint optimization problem of the RIS-FAS system in a field-driven manner. As defined in~(\ref{eq:fas_sampling}) and~(\ref{eq:fas_ris_received}), the received signal at the $m$-th FA can be expressed as
\begin{align}\label{eq:rx_signal_interference}
y_m = \sqrt{P}\, E_{k}(\xi_m)
+ \sum_{j \neq k} \sqrt{P}\, E_{j}(\xi_m)
+ \sigma^2,
\mathbf{\xi}_m \in \mathcal{S}_t,
\end{align}
where $E_{k}(\xi_m)$ denotes the desired-user field distribution defined in~(\ref{eq:fas_ris_field}),
$E_{j}(\mathbf{\xi}_m)$ represents the interference field contributed by user~$j$.

Based on the previously established 3DGRF and the received signal, the instantaneous angular power spectrum of the system is defined as:
\begin{equation}
\Phi(\overline{\xi},\boldsymbol{\Theta}) =
\frac{1}{M}\sum_{m=1}^{M}
\left|E_{k}(\mathbf{\xi}_m)\right|^2,
\label{eq:Prad}
\end{equation}
which directly characterizes the spatial distribution of the radiation field's energy. 

The interference power spectrum is given by
\begin{equation}
\Phi_{j}(\overline{\mathbf{\xi}},\boldsymbol{\Theta}) =
\frac{1}{M}\sum_{m=1}^{M}
\sum_{j \neq k}
\left|E_{j}(\mathbf{\xi}_m)\right|^2.
\label{eq:Pint}
\end{equation}

The overall received SINR can then be expressed as

\begin{equation}
\Gamma(\overline{\mathbf{\xi}},\boldsymbol{\Theta})
=\frac{P\, \Phi(\overline{\mathbf{\xi}},\boldsymbol{\Theta})}
{P\Phi_{j}(\overline{\mathbf{\xi}},\boldsymbol{\Theta})+\sigma^2}.
\label{eq:SINR}
\end{equation}

Accordingly, the optimization objective $R$ can be rewritten as \eqref{eq:rate}.
The joint optimization problem is then formulated as:

\begin{equation}
\begin{aligned}
\max_{\overline{\mathbf{s}},\boldsymbol{\Theta}} \quad
& \log_2\!\left(1 + \frac{
P\, \frac{1}{M}\sum_{m=1}^M \big|\Theta_{km}\big|^2
}{
P\frac{1}{M}\sum_{m=1}^M \sum_{j\neq k} \big|\Theta_{jm}\big|^2 + \sigma^2
}\right) \\
\text{s.t.}\quad
& \mathbf{\xi}_m \in S_t,\ \|\mathbf{\xi}_m-\mathbf{\xi}_v\|_2 \ge D\ (m\neq v), \\
& \theta_n \in \{0,\tfrac{2\pi}{L_c},\ldots,2\pi(1-\tfrac{1}{L_c})\},
\end{aligned}
\label{eq:joint_opt}
\end{equation}
where the inner summation term is simplified as
\[
\Theta_{jm} = \sum_{i=1}^J A_i^{(j)} e^{j\theta_i}
\exp\!\left(-\tfrac{1}{2}(\mathbf{\xi}_m-\mathbf{q}_i)^{\mathrm{T}}
\boldsymbol{\Sigma}_i^{-1}(\mathbf{\xi}_m-\mathbf{q}_i)\right)
\]
where $A_i^{(j)}$ denotes the amplitude coefficient of the $i$-th Gaussians for FA $m$'s signal path.

Here, $L_c$ represents the phase quantization level. This problem is a typical mixed non-convex optimization problem, where $\overline{\mathbf{\xi}}$ is a continuous variable and $\boldsymbol{\Theta}$ is discrete. Moreover, the strong coupling between these variables in the objective function makes direct optimization intractable.
To address this issue, a FAO method is proposed. The core idea is to iteratively update the FAS positions and RIS phases by leveraging the power-spectrum gradients reconstructed from the 3DGRF, rather than relying on explicit channel estimation matrices.

\subsection{FAS Position Optimization}

When the RIS reflection matrix $\boldsymbol{\Theta}$ is fixed, the subproblem of optimizing the FAS positions can be formulated as:
\begin{equation}
\max_{\overline{\mathbf{\xi}}} \
\Phi(\overline{\mathbf{\xi}};\boldsymbol{\Theta})
\quad
\text{s.t. } \mathbf{\xi}_m\in S_\xi,\ \|\mathbf{\xi}_m-\mathbf{\xi}_v\|_2\ge D.
\label{eq:FAS_opt_problem}
\end{equation}

Since the objective function $\Phi(\overline{\mathbf{\xi}};\boldsymbol{\Theta})$ is non-convex with respect to $\overline{\mathbf{\xi}}$, we adopt the \emph{successive convex approximation} (SCA) technique to iteratively linearize the objective around the current position $\overline{\mathbf{\xi}}^{(v)}$~\cite{SCA}.
Specifically, the first-order Taylor expansion is employed to construct a locally convex surrogate function:
\begin{equation}
\widetilde{\Phi}(\overline{\mathbf{\xi}})
= \Phi(\overline{\mathbf{\xi}}^{(v)})
+ \sum_{m=1}^{M}
\nabla_{\mathbf{\xi}_m} \Phi(\overline{\mathbf{\xi}}^{(v)})^{\!\top}
(\mathbf{\xi}_m - \mathbf{\xi}_m^{(v)}),
\label{eq:sca_approx}
\end{equation}
where the gradient $\nabla_{\mathbf{\xi}_m} \Phi$ is given by
\begin{equation}
\nabla_{\mathbf{\xi}_m} \Phi
= \sum_{n=1}^{N}
\Re\!\left\{
E_{k}^*(\mathbf{\xi}_m)
\frac{\partial E_{k}(\mathbf{\xi}_m)}{\partial \mathbf{\xi}_m}
\right\}.
\label{eq:FAS_grad}
\end{equation}

The convexified subproblem in the $(q+1)$-th iteration can thus be written as:
\begin{equation}
\max_{\overline{\mathbf{\xi}}}\
\widetilde{\Phi}(\overline{\mathbf{\xi}})
\quad
\text{s.t. } \mathbf{\xi}_m \in S_\xi,\
\|\mathbf{\xi}_m - \mathbf{\xi}_v\|_2 \ge D.
\label{eq:FAS_sca_problem}
\end{equation}

This convex optimization can be efficiently solved using standard solvers.
After convergence, the antenna positions are updated as
\begin{equation}
\mathbf{\xi}_m^{(v+1)} = \mathbf{\xi}_m^{(v)} + \vartheta_\xi
(\mathbf{\xi}_m^{*} - \mathbf{\xi}_m^{(v)}),
\label{eq:FAS_update_sca}
\end{equation}
where $\vartheta_\xi$ is the adaptive step-size ensuring monotonic improvement of the objective.  
Through successive convex refinements, the FAS positions gradually converge to a locally optimal configuration that maximizes the radiation-field power spectrum.

The overall FAS position optimization is summarized in \textbf{Algorithm~\ref{alg:FAS_opt}}.

\begin{algorithm}[t]
\caption{FAS Position Optimization under Fixed RIS}
\label{alg:FAS_opt}
\begin{algorithmic}
\REQUIRE Fixed RIS $\boldsymbol{\Theta}$, initial FAS positions $\overline{\mathbf{\xi}}^{(0)}$, step size $\vartheta_\xi$, threshold $\epsilon_\xi$, max iterations $Q_\xi$
\ENSURE Optimized FAS positions $\overline{\mathbf{\xi}}^{*}$

\FOR{$v = 0$ \textbf{to} $Q_\xi$}
    \STATE Compute gradient $\nabla_{\mathbf{\xi}_m} \Phi$ as in \eqref{eq:FAS_grad}
    \STATE Update $\mathbf{\xi}_m$ using \eqref{eq:FAS_update_sca} and project onto $S_\xi$
    \STATE Compute $\Phi^{(v+1)}$ using \eqref{eq:Prad}
    \IF{$|\Phi^{(v+1)} - \Phi^{(v)}| < \epsilon_\xi$}
        \STATE \textbf{break}
    \ENDIF
\ENDFOR

\STATE \textbf{return} $\overline{\mathbf{\xi}}^{*} = \overline{\mathbf{\xi}}^{(v+1)}$
\end{algorithmic}
\end{algorithm}

\subsection{RIS Phase Configuration Optimization}
When the FAS positions $\overline{\mathbf{s}}$ are fixed, the optimization of the RIS phase shifts can be expressed as:
\begin{equation}
\max_{\boldsymbol{\Theta}} \
\Phi(\overline{\mathbf{\xi}};\boldsymbol{\Theta})
\quad
\text{s.t. } \theta_n \in \mathcal{C},
\label{eq:RIS_opt_problem_GA}
\end{equation}
where $\mathcal{C}$ denotes the discrete codebook set \cite{codebook} of phase quantization levels.  

To efficiently explore the non-convex discrete search space, we employ a \emph{genetic algorithm} (GA)~\cite{9366346}.  
Each individual in the GA population represents a candidate RIS phase vector 
$\boldsymbol{\theta} = [\theta_1, \ldots, \theta_N]$, 
and its fitness function is defined as the corresponding radiation power:
\begin{equation}
F(\boldsymbol{\theta}) = 
\Phi(\overline{\mathbf{\xi}};\boldsymbol{\Theta}(\boldsymbol{\theta})).
\label{eq:fitness_func}
\end{equation}

For each RIS element $n$, the marginal power contribution of a candidate phase $\theta$ is defined as:
\begin{equation}
\Delta P_n(\theta) = 
F([\theta_1, \ldots, \theta_{n-1}, \theta, \theta_{n+1}, \ldots, \theta_N])
- F(\boldsymbol{\theta}).
\label{eq:deltaP}
\end{equation}

During the optimization process, the GA evolves the population through three key operations:  
selection, crossover, and mutation. In each generation, individuals with higher fitness values are preferentially selected to preserve promising phase configurations. The crossover operation then combines partial phase vectors from selected pairs of individuals to generate new offspring, enabling information exchange among high-quality solutions. To prevent premature convergence and maintain diversity, the mutation step randomly perturbs several phase entries with a small probability. Over $Q_{\text{GA}}$ generations of evolution, the population gradually converges to an optimal or near-optimal RIS phase configuration.

After $Q_{\text{GA}}$ generations, the best individual $\boldsymbol{\theta}^*$ is adopted as the optimized RIS phase configuration:
\begin{equation}
\boldsymbol{\Theta}^* = 
\mathrm{diag}\{ e^{j\theta_1^*}, e^{j\theta_2^*}, \ldots, e^{j\theta_N^*} \}.
\label{eq:RIS_GA_opt}
\end{equation}

The GA-based discrete optimization effectively avoids local minima and enables near-global search without explicit gradient computation.

The overall Discrete Optimization of RIS Phases is summarized in \textbf{Algorithm~\ref{alg:RIS_opt}}.

\begin{algorithm}[t]
\caption{RIS Phase Optimization under Fixed FAS}
\label{alg:RIS_opt}
\begin{algorithmic}
\REQUIRE Fixed FAS $\overline{\mathbf{s}}$, initial RIS $\boldsymbol{\Theta}^{(0)}$, codebook $\mathcal{C}$, threshold $\epsilon_r$, max iterations $Q_r$
\ENSURE Optimized RIS phases $\boldsymbol{\Theta}^{*}$

\FOR{$v = 0$ \textbf{to} $Q_r$}
    \FOR{$n = 1$ \textbf{to} $N$}
        \STATE Compute $\Delta P_n(\theta_n)$ as in \eqref{eq:deltaP}  
        \STATE Select $\theta_n^{(v+1)} = \arg\max_{\theta \in \mathcal{C}} \Delta P_n(\theta)$
    \ENDFOR
    \STATE Form updated RIS $\boldsymbol{\Theta}^{(v+1)} = \mathrm{diag}(e^{j\theta_1^{(v+1)}},\ldots,e^{j\theta_N^{(v+1)}})$
    \STATE Compute $\Phi^{(v+1)}$ using \eqref{eq:Prad}
    \IF{$|\Phi^{(v+1)} - \Phi^{(v)}| < \epsilon_r$}
        \STATE \textbf{break}
    \ENDIF
\ENDFOR

\STATE \textbf{return} $\boldsymbol{\Theta}^{*} = \boldsymbol{\Theta}^{(v+1)}$
\end{algorithmic}
\end{algorithm}

\subsection{Field-Driven Alternating Optimization Framework}

\begin{algorithm}[t]
\caption{Field-Driven Optimization using 3DGRF}
\label{alg:FDAO}
\begin{algorithmic}
\REQUIRE Initial FAS $\overline{\mathbf{\xi}}^{(0)}$, RIS $\boldsymbol{\Theta}^{(0)}$, max iterations $Q_{\max}$, threshold $\epsilon$
\ENSURE Optimized $\overline{\mathbf{\xi}}^{*}$, $\boldsymbol{\Theta}^{*}$

\FOR{$v = 0$ \textbf{to} $Q_{\max}$}
    \STATE \textbf{FAS Update:} Fix $\boldsymbol{\Theta}^{(v)}$, update $\overline{\mathbf{\xi}}^{(v+1)}$ using \eqref{eq:FAS_grad}--\eqref{eq:FAS_update_sca}
    \STATE \textbf{RIS Update:} Fix $\overline{\mathbf{\xi}}^{(v+1)}$, update $\boldsymbol{\Theta}^{(v+1)}$ using \eqref{eq:deltaP} and \eqref{eq:RIS_GA_opt}
    \STATE Compute $\Phi^{(v+1)}$ using \eqref{eq:Prad}
    \IF{$|\Phi^{(v+1)} - \Phi^{(v)}| < \epsilon$}
        \STATE \textbf{break}
    \ENDIF
\ENDFOR

\STATE \textbf{return} $\overline{\mathbf{\xi}}^{*} = \overline{\mathbf{\xi}}^{(v+1)},\ \boldsymbol{\Theta}^{*} = \boldsymbol{\Theta}^{(v+1)}$
\end{algorithmic}
\end{algorithm}

The overall optimization framework alternates between the SCA-based continuous optimization of FAS positions and the GA-based discrete optimization of RIS phases, as summarized in \textbf{Algorithm~\ref{alg:FDAO}}.  

Each iteration guarantees a non-decreasing radiation-field power, and since the overall objective is upper-bounded, the algorithm converges to a locally optimal field configuration.  
The optimization process operates directly on the radiation-field representation, enabling adaptive control of both FAS positions and RIS phases without relying on explicit channel estimation or convex relaxations.  

In terms of computational complexity, the SCA-based FAS position update primarily involves gradient-based matrix-vector operations that scale linearly with the number of FAS elements $M$.  
Meanwhile, the GA-based RIS phase update requires evaluating the fitness of $N$ reflecting elements with $L_c$ bits.  
Therefore, the overall per-iteration complexity of the proposed framework can be expressed as
\begin{equation}
\mathcal{O}(M + N^{|L_c|}),
\end{equation}
which reflects the linear scalability of the proposed field-driven optimization with respect to the system dimensions.  
This ensures that the framework can efficiently adapt to large-scale RIS-assisted deployments in real time.

\subsection{Complexity Analysis}

The computational complexity of the proposed alternating optimization framework mainly arises from two stages: the SCA-based FAS position update and the GA-based RIS phase optimization.  
In each outer iteration, denoted by $Q$, these two modules are executed sequentially to refine the field configuration.

In the FAS optimization stage, the dominant operations come from evaluating the radiation-field gradient and solving the convexified subproblem in \eqref{eq:FAS_sca_problem}.  
For each antenna element, the gradient $\nabla_{\mathbf{\xi}_m} \Phi$ in \eqref{eq:FAS_grad} requires accumulating the contributions of all $N$ RIS reflecting elements, which results in a computational cost of $\mathcal{O}(MN)$.
The subsequent projection and position update in \eqref{eq:FAS_update_sca} involve only element-wise operations and therefore add negligible overhead.  
Hence, the overall complexity of the FAS update per iteration can be approximated as
\begin{equation}
C_{\text{FAS}} \approx \mathcal{O}(MN).
\end{equation}

In the RIS optimization stage, the genetic algorithm evaluates the fitness function defined in \eqref{eq:fitness_func} for a population of $G_{\text{pop}}$ individuals over $G_{\text{iter}}$ generations.  
Each evaluation involves computing the total radiation power contributed by $N$ reflecting elements, leading to a per-generation cost of $\mathcal{O}(G_{\text{pop}}N)$.  
Therefore, the total computational complexity of the GA-based phase optimization can be expressed as
\begin{equation}
C_{\text{RIS}} \approx \mathcal{O}(G_{\text{pop}}G_{\text{iter}}N).
\end{equation}

Combining both stages, the overall complexity of one outer iteration of the proposed framework is given by
\begin{equation}
C_{\text{iter}} = C_{\text{FAS}} + C_{\text{RIS}} 
\approx \mathcal{O}(MN + G_{\text{pop}}G_{\text{iter}}N).
\end{equation}
After $Q$ alternating iterations, the total computational complexity can thus be summarized as
\begin{equation}
\mathcal{O}\!\left(Q(MN + G_{\text{pop}}G_{\text{iter}}N)\right).
\label{eq:complexity_new}
\end{equation}

Since both the gradient computations and fitness evaluations are highly parallelizable, the proposed field-driven optimization can be efficiently implemented on GPUs.  
In practice, the number of antennas $M$ and reflecting elements $N$ are moderate, while $G_{\text{pop}}$ and $G_{\text{iter}}$ are typically small to ensure real-time convergence.  
As a result, the proposed framework enables fast reconfiguration, well-suited to dynamic 6G communication environments.

\begin{figure*}
    \centering
    \includegraphics[width=\linewidth]{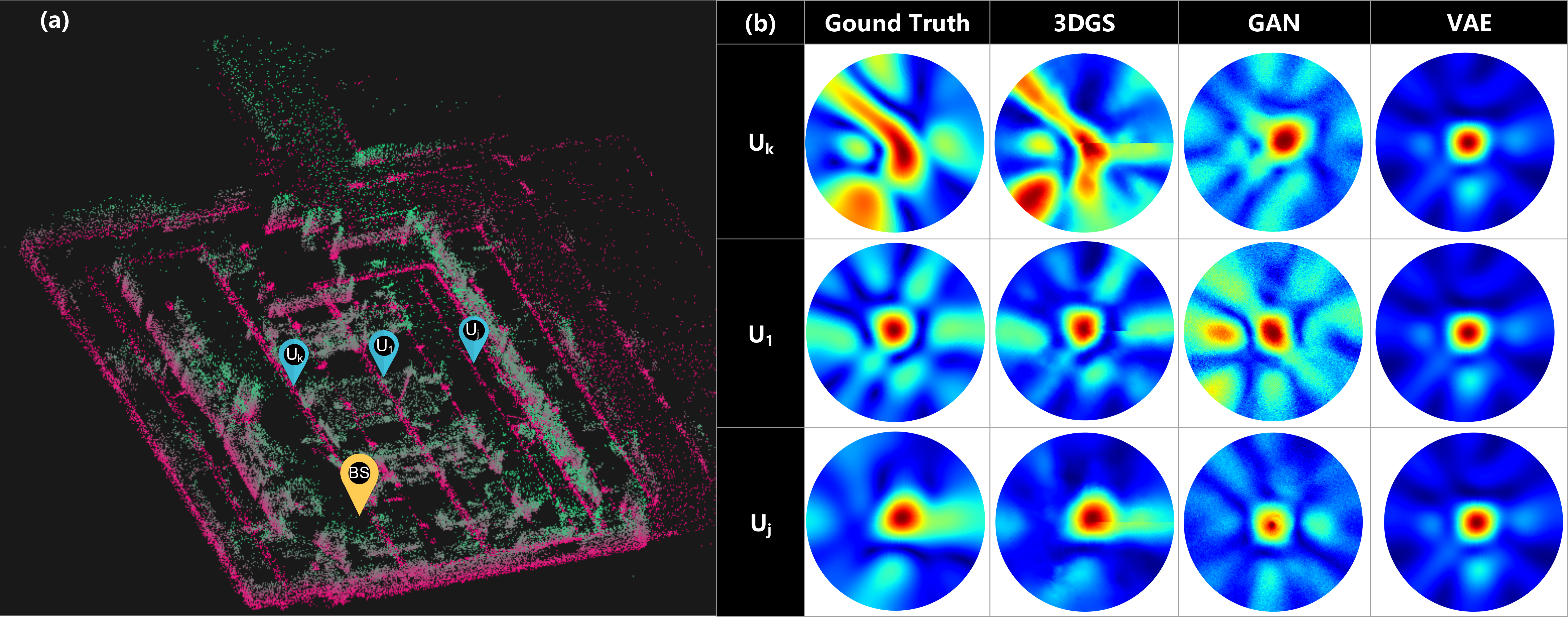}
    \caption{(a)Laboratory environment and LiDAR point-cloud representation for radiation-field reconstruction, where the base station (BS) and multiple users $U_k,U_1,U_j$ are marked. 3D LiDAR point clouds of the laboratory environment were used to initialize the Gaussian primitives for the proposed 3DGS-based field model. The overall experimental scene, where the transmitter (TX) can be placed at arbitrary positions within the environment, and the receiver (RX) equipped with a $4\times4$ antenna array is fixed at the corner of the room.}
    \label{fig:field_slice}
\end{figure*}

\section{Numerical Results and Discussion}\label{s5}
\label{sec:experiment}

\subsection{Implementation for 3DGRF Reconstruction}
3DGRF experiments are conducted on a workstation equipped with an NVIDIA RTX~4090~GPU with CUDA kernels, an Intel(R) Silver(R) 4410Y CPU, and 128~GB RAM. The 3DGRF reconstruction and optimization are implemented using PyTorch~2.3.1 and CUDA~12.2. All GPU operations, including Gaussian rendering and field-projection operations, are parallelized to achieve real-time training and inference.

We use an open-source dataset provided in \cite{12} to evaluate the 3DGRF framework. The physical environment for radiation-field reconstruction is shown in Fig.~\ref{fig:field_slice}. A 3D LiDAR sensor is employed to capture the surrounding geometry, including walls, tables, and reflectors. The 3D LiDAR point clouds of the environment are shown in Fig.~\ref{fig:field_slice}(a). The receiver (RX) is equipped with a $4\times4$ antenna array, while the transmitter (TX) continuously sends messages. The RX position is fixed, and the TX position is systematically varied within the laboratory. Each data sample comprises a TX coordinate and the corresponding measured spatial spectrum at the RX. The dataset includes 6,000 samples, with 80\% for training and 20\% for testing. The obtained 3D point cloud provides a dense geometric representation of the laboratory environment, containing approximately $1.5\times10^6$~points. These points are preprocessed by downsampling and normal estimation to initialize the spatial distribution of Gaussian primitives. Each LiDAR point corresponds to an initial Gaussian center $\boldsymbol{\mu}_q$, whose covariance $\boldsymbol{\Sigma}_q$ is determined by local point density and surface curvature, thereby encoding geometric smoothness into the initial radiation-field model. The position and orientation of the Gaussian primitives are updated adaptively during network optimization.

\subsubsection{Scene Representation Network}

The scene representation network (SRN) consists of two MLPs. The first MLP encodes the spatial geometry of each Gaussian primitive and predicts amplitude attenuation $\mu(\mathbf{q})$, while the second predicts phase offset $\delta(\mathbf{q})$ conditioned on the transmitter position. Each MLP contains four hidden layers with 256 neurons and uses LeakyReLU activation. The network is trained with Adam optimizer using an initial learning rate of $5\times10^{-4}$, decayed exponentially by~0.95 every~20~epochs.

The loss function jointly measures amplitude fidelity and structural similarity between the reconstructed and ground-truth fields, as shown in (\ref{eq:srn_loss}).

\subsubsection{Radiation Field Reconstruction Baselines}
To evaluate the fidelity of the proposed 3DGRF reconstruction, we compare it against three representative baselines:

\begin{itemize}
\item \textbf{Generative Adversarial Network (GAN)\cite{GAN}:} A data-driven benchmark that directly learns the nonlinear mapping from transmitter or RIS–FAS configurations to the corresponding power or radiation field distributions. The generator captures global spatial correlations through adversarial training against a discriminator.
\item \textbf{Variational Autoencoder (VAE)\cite{VAE}:} A probabilistic generative model that embeds the field distribution into a latent Gaussian manifold. By sampling from the learned latent space, the VAE reconstructs continuous radiation maps with smooth energy variations. 
\item \textbf{Probabilistic Channel Model (PCM):} A physics-inspired baseline that characterizes the wireless propagation using stochastic statistical parameters such as path loss, Rician fading, and spatial correlation matrices. Optimization is performed in the channel domain by iteratively updating RIS phases and FAS positions based on estimated CSI.
\end{itemize}

\subsection{System Parameters for Communications}
\begin{table}[t]
\centering
\caption{Simulation Parameters}
\label{tab:sim_params}
\begin{tabular}{lcc}
\toprule
\textbf{Parameter} & \textbf{Symbol} & \textbf{Value} \\
\midrule
Number of RIS elements & $N$ & 64 \\
Phase quantization level & $L_c$ & 4~(2-bit) \\
Number of FAS antennas & $M$ & 16 \\
Minimum spacing & $W$ & $\lambda$ \\
Transmit power & $P$ & 10~dBm \\
Noise power & $\sigma^2$ & $-90$~dBm \\
Convergence threshold & $\varepsilon$ & $10^{-4}$ \\
Learning rate & $lr$ & $5\times10^{-4}$ \\
\bottomrule
\end{tabular}
\end{table}

The main simulation parameters for the communication environment are summarized in Table~\ref{tab:sim_params}. 

To comprehensively evaluate the proposed FAO framework, we compare it with two categories of reference schemes: (i) \textit{optimization methods} that evaluate the effectiveness of the proposed algorithm, and (ii) \textit{system baselines} that reveal the importance of FAS and RIS for achieving the rate maximum.

\subsubsection{Optimization Methods}
\begin{itemize}
\item \textbf{GD\cite{GD}:} A classical continuous-phase optimization method that updates each RIS reflection coefficient through projected gradient descent under the unit-modulus constraint. After convergence, the optimized continuous phases are uniformly quantized to a 2-bit codebook \cite{2bit}. This approach reflects conventional differentiable optimization performed in the channel domain.
\item \textbf{ADMM\cite{ADMM}:} A physics-consistent convex-relaxation framework that decomposes the coupled FAS–RIS joint optimization into several tractable subproblems with separable constraints. Each subproblem can be solved efficiently with guaranteed convergence to a locally optimal stationary point.
\end{itemize}
Both serve as algorithmic baselines that approximate the physical-space optimization process but remain channel-dependent, providing insight into the advantages of direct field-domain optimization achieved by 3DGRF-driven FAO.

\subsubsection{System Baselines}
\begin{itemize}
\item \textbf{w/o RIS:} The RIS module is deactivated, and only the direct BS–user transmission is considered. This baseline represents a conventional FAS-assisted system without any reflective enhancement and serves to quantify the net gain introduced by RIS deployment.
\item \textbf{Random:}The RIS is activated with random discrete phase assignments selected from a 2-bit quantization. This case evaluates the effect of random scattering without optimization.
\item \textbf{FPA:} The fluid antenna system operates with fixed antenna positions, while the RIS is fully optimized. This baseline removes the mobility degree of freedom of FAS, revealing the pure benefit of spatial adaptability brought by antenna movement.
\end{itemize}
These baselines jointly quantify the impact of each functional component—RIS reconfiguration and FAS movement—on overall system performance, enabling a clear demonstration of the effectiveness of the proposed 3DGRF-driven optimization framework.

\subsection{Radiation Field Modeling Performance}

Fig. \ref{fig:field_slice}(b) compares the angular-domain power spectra reconstructed by different radiation-field modeling methods, including the proposed 3DGS, GAN, and VAE. The ground-truth spectrum is obtained from real-world ray-tracing measurements calibrated by site-specific channel sounding, serving as a reference for field reconstruction. As observed, the 3DGS model accurately reproduces both the global radiation pattern and the fine interference fringes of the measured field, exhibiting almost perfect alignment with the ground truth. In contrast, the GAN and VAE baselines exhibit clear spatial blurring and phase inconsistencies, particularly around high-energy lobes and null regions.
These results verify that the explicit Gaussian-primitive representation in 3DGS effectively preserves sub-wavelength spatial energy variations and phase continuity, leading to physically faithful field modeling.
Such high-fidelity reconstruction provides a reliable foundation for subsequent field-driven RIS–FAS optimization, ensuring that the learned radiation field can faithfully guide configuration decisions in real propagation environments.
\subsection{Visualization of Radiation Field Spectrum}
\begin{figure}[t]
    \centering
    \begin{subfigure}{0.45\linewidth}
        \includegraphics[width=\textwidth]{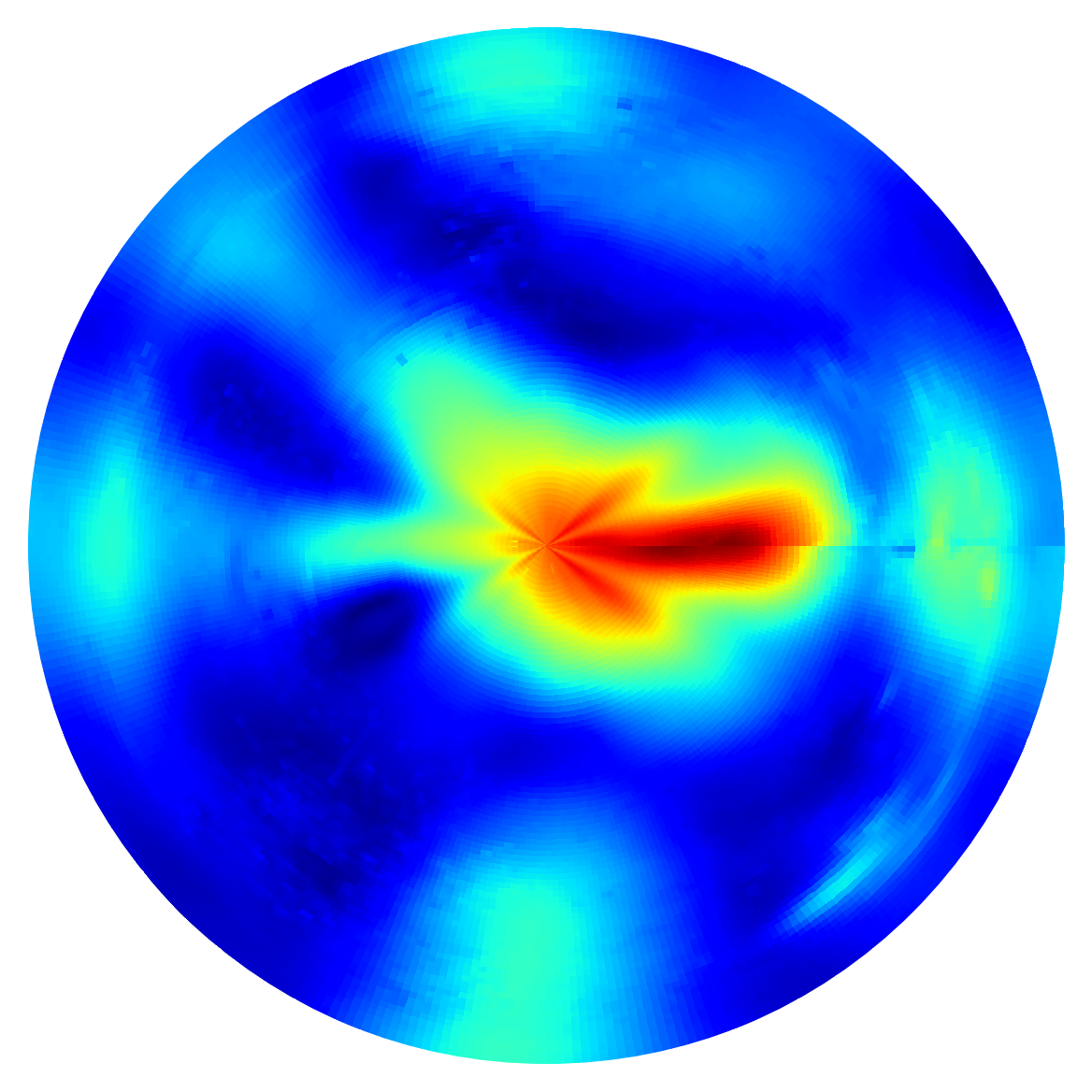}
        \caption{}

    \end{subfigure}
    \begin{subfigure}{0.45\linewidth}
        \includegraphics[width=\textwidth]{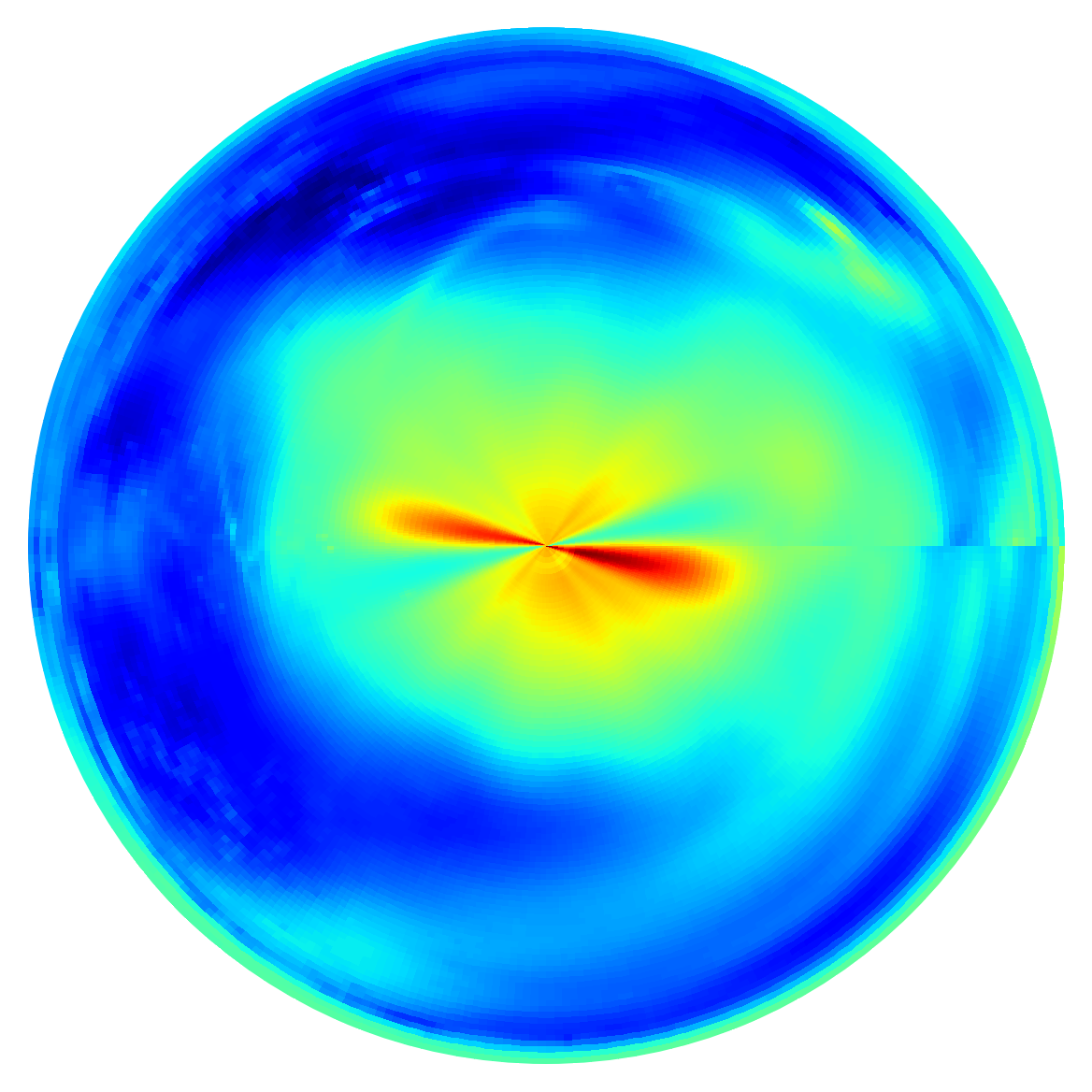}
        \caption{}
    \end{subfigure}
    \caption{(a) Synthesized spatial power spectrum of the RIS–FAS system before optimization, where the integration of RIS reflection and FAS reception forms a preliminary directional pattern. (b) Spectrum after field-driven optimization, showing stronger beam focusing and reduced sidelobes, which demonstrates the effectiveness of the proposed joint configuration of RIS phase and FAS position.}
    \label{fig:spectrum_plus}
\end{figure}

In Fig. \ref{fig:spectrum_plus}(a), the spatial spectrum corresponds to the initial configuration after integrating the FAS and RIS modules. The radiation energy becomes directionally enhanced compared with the case in Fig. \ref{fig:field_slice}(b), indicating that the joint reflection–reception structure already forms a preliminary beam pattern toward the dominant propagation direction.
After performing the proposed field-driven alternating optimization, Fig. \ref{fig:spectrum_plus}(b) shows that the main lobe becomes significantly narrower and the sidelobes are largely suppressed, demonstrating a well-focused beamforming effect. This observation confirms that the optimized RIS phase and FAS position jointly steer the electromagnetic field to the desired spatial region, achieving stronger energy concentration and more efficient spatial utilization. The result verifies the physical interpretability of the 3DGRF model, which directly translates field-domain optimization into observable beam enhancement.

\subsection{Transmission Rate Comparison}
\begin{figure}[!t]
    \centering
    \includegraphics[width=0.8\linewidth]{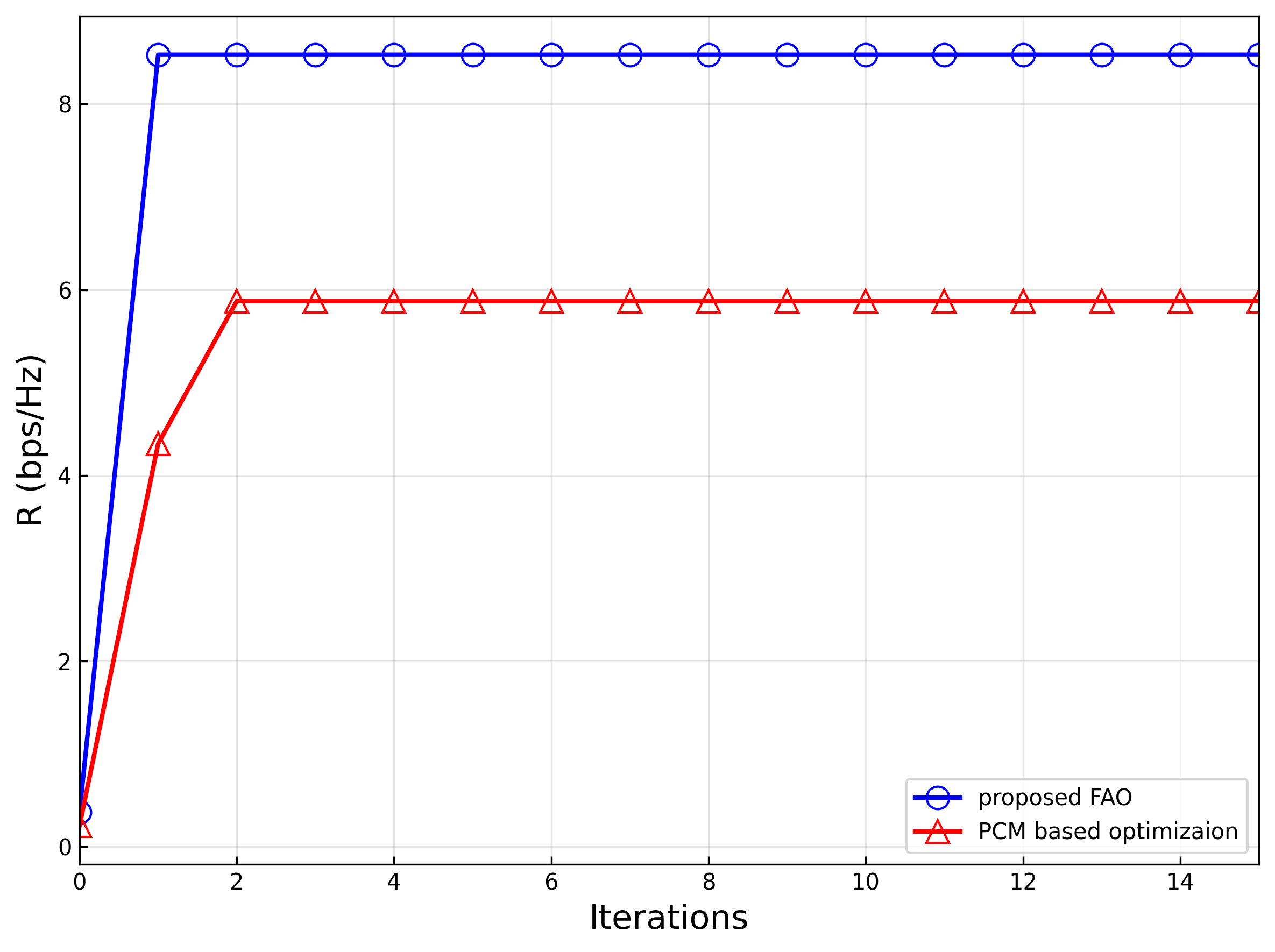}
    \caption{Minimum achievable rate $R$ comparison between the \textit{proposed} field-driven optimization and the \textit{traditional} optimization.}
    \label{fig:Tovs3dgs}
\end{figure}

Fig.~\ref{fig:Tovs3dgs} illustrates the evolution of the minimum achievable rate versus iteration number for the proposed field-driven optimization and the conventional optimization under identical system settings.
As the iteration count increases, the achievable rates of both methods gradually stabilize, indicating that each algorithm reaches convergence. Notably, the proposed approach converges within fewer iterations and attains a markedly higher steady-state rate, demonstrating both faster convergence and superior performance. This performance gain originates from the 3DGRF. By leveraging the learned power spectrum, the optimizer can rapidly align the transceiver configuration with the actual energy distribution in the environment, thereby improving spectral efficiency and robustness in dynamic or fast-fading conditions. These results verify that the explicit Gaussian field modeling bridges the gap between physical-space representation and communication optimization. It achieves fast, stable convergence without requiring channel estimation, while maintaining low computational complexity and pilot overhead.

\subsection{Average Delay Results}

\begin{figure}[t]
    \centering
    \includegraphics[width=\linewidth]{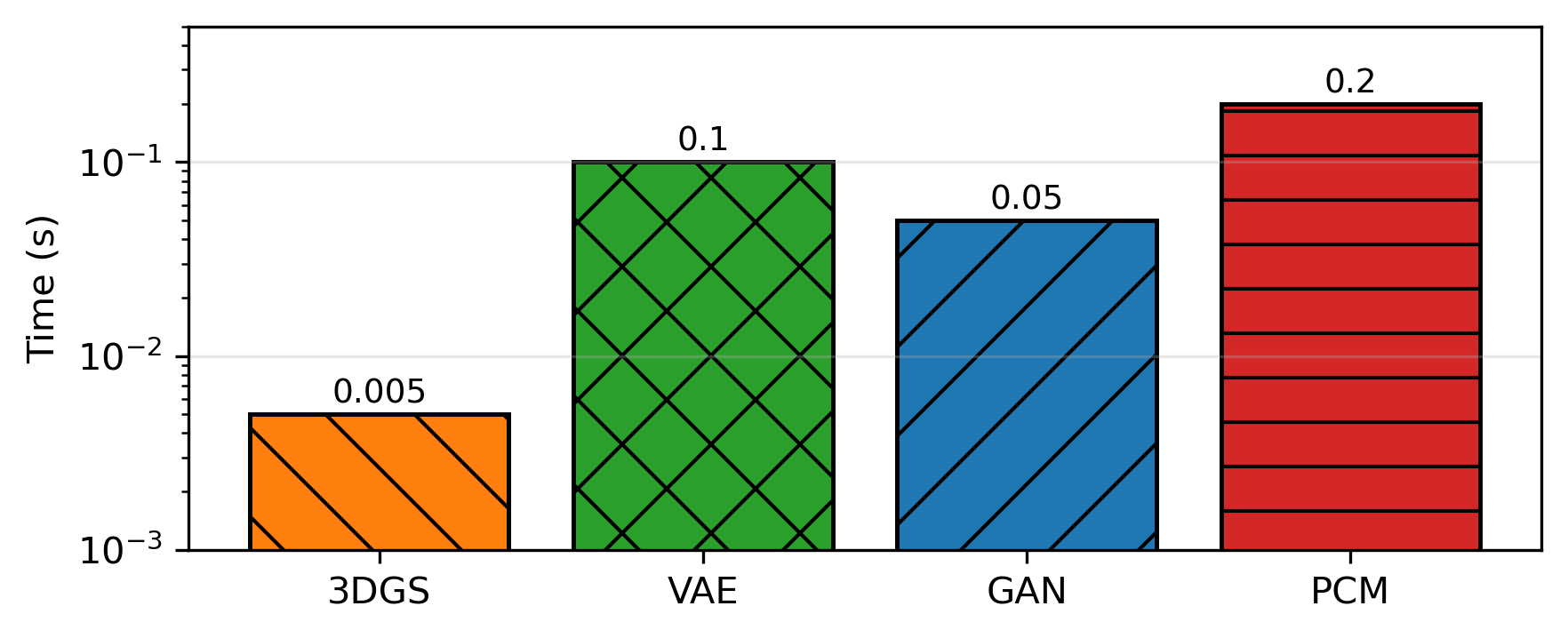}
    \caption{Average delay versus different channel modeling methods.}
    \label{fig:timeCompa}
\end{figure}

Fig.~\ref{fig:timeCompa} compares the average processing latency of four optimization frameworks that employ different channel modeling paradigms: the proposed 3DGS model, VAE, GAN, and PCM–based optimization. All methods share the same inference and optimization pipeline, while differing only in how the underlying propagation environment is represented and updated. As shown, the proposed 3DGS-based method achieves the lowest latency, with an average computation time of only $5\times10^{-3}$~s, which is about 10$\times$ faster than the probabilistic model–based traditional optimization ($0.2$~s) and significantly more efficient than the data-driven baselines VAE ($0.1$~s) and GAN ($0.05$~s).

This demonstrates that the 3DGS framework not only converges faster but also achieves real-time adaptability with minimal computational overhead. The gain arises from the 3DGS model's radiation-field reconstruction mechanism, which directly infers the receiver-side radio-frequency field from the transmitter geometry, enabling continuous field representation without repeated channel estimation or iterative matrix inversion. In contrast, the VAE and GAN baselines can also infer the radiation field but rely on deep encoder–decoder architectures with large parameter spaces, resulting in longer inference times and higher GPU memory usage. The probabilistic channel model further incurs heavy iterative computation due to stochastic channel sampling and expectation-based optimization. \emph{Results show that the proposed 3DGS model provides a lightweight and physically interpretable alternative that supports low-latency, real-time RIS–FAS optimization.}

\subsection{Achievable Sum-Rate Results}
\begin{figure}[t]
    \centering
    \includegraphics[width=0.8\linewidth]{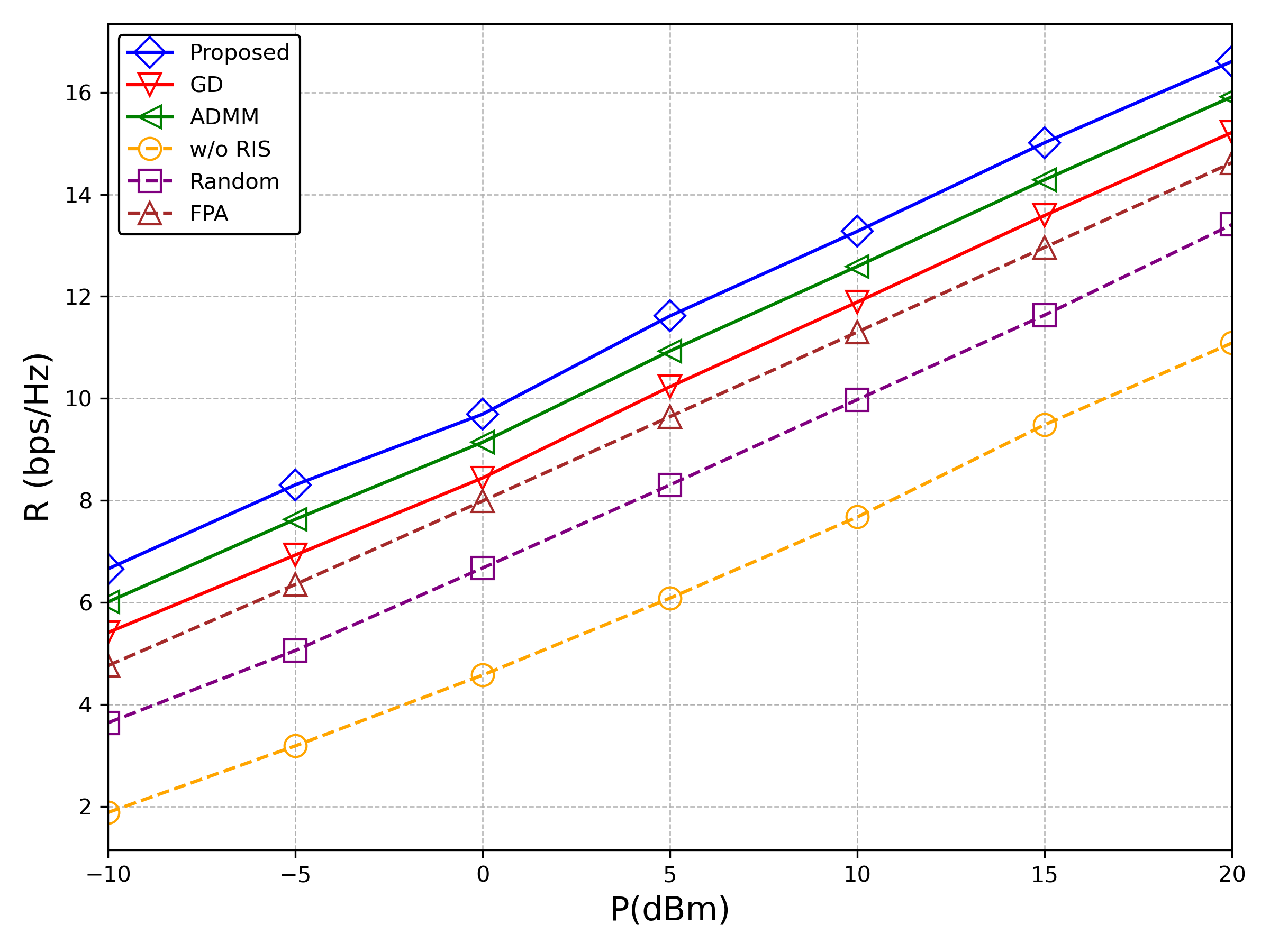}
    \caption{Transmit power versus minimum achievable rate $R$.}
    \label{fig:Power}
\end{figure}

Fig.~\ref{fig:Power} illustrates the achievable rate performance versus transmit power $P\in[-10,20]dBm$ under identical RIS–FAS configurations. 
As expected, all curves exhibit a monotonic increase with transmit power, confirming the theoretical SNR dependence of achievable rate. The proposed FAO consistently achieves the highest rate across the entire power range. This superior performance stems from the 3DGRF modeling, which provides a differentiable, physically grounded description of both the desired signal focusing and the reflected interference field. By learning the continuous radiation distribution, the FAO jointly optimizes FAS positions and RIS phases to concentrate energy toward the desired user while simultaneously suppressing undesired reflections from interfering users. This field-domain coordination enables coherent power aggregation and adaptive interference control, thereby maximizing the minimum achievable rate in multiuser conditions. The GD and ADMM schemes follow similar upward trends but reach lower saturation levels. Their channel-domain formulations rely on discrete CSI and iterative convex updates, which approximate the alignment of reflection phases and cannot fully mitigate inter-user interference in the continuous radiation field. The w/o RIS baseline lacks reflective diversity, resulting in poor energy utilization. The Random configuration suffers from uncontrolled scattering and constructive–destructive phase randomness, yielding severe interference fluctuations. The FPA case fixes antenna positions, thereby losing the spatial adaptability that enables FAS to exploit local radiation peaks, resulting in a moderate but limited improvement.
\emph{The consistent superiority of the FAO curve across all power levels proves that field-aware optimization provides better energy focusing and robustness than channel-driven schemes, reinforcing the efficiency of 3DGRF-based field modeling.}

\begin{figure}[t]
    \centering
    \includegraphics[width=0.8\linewidth]{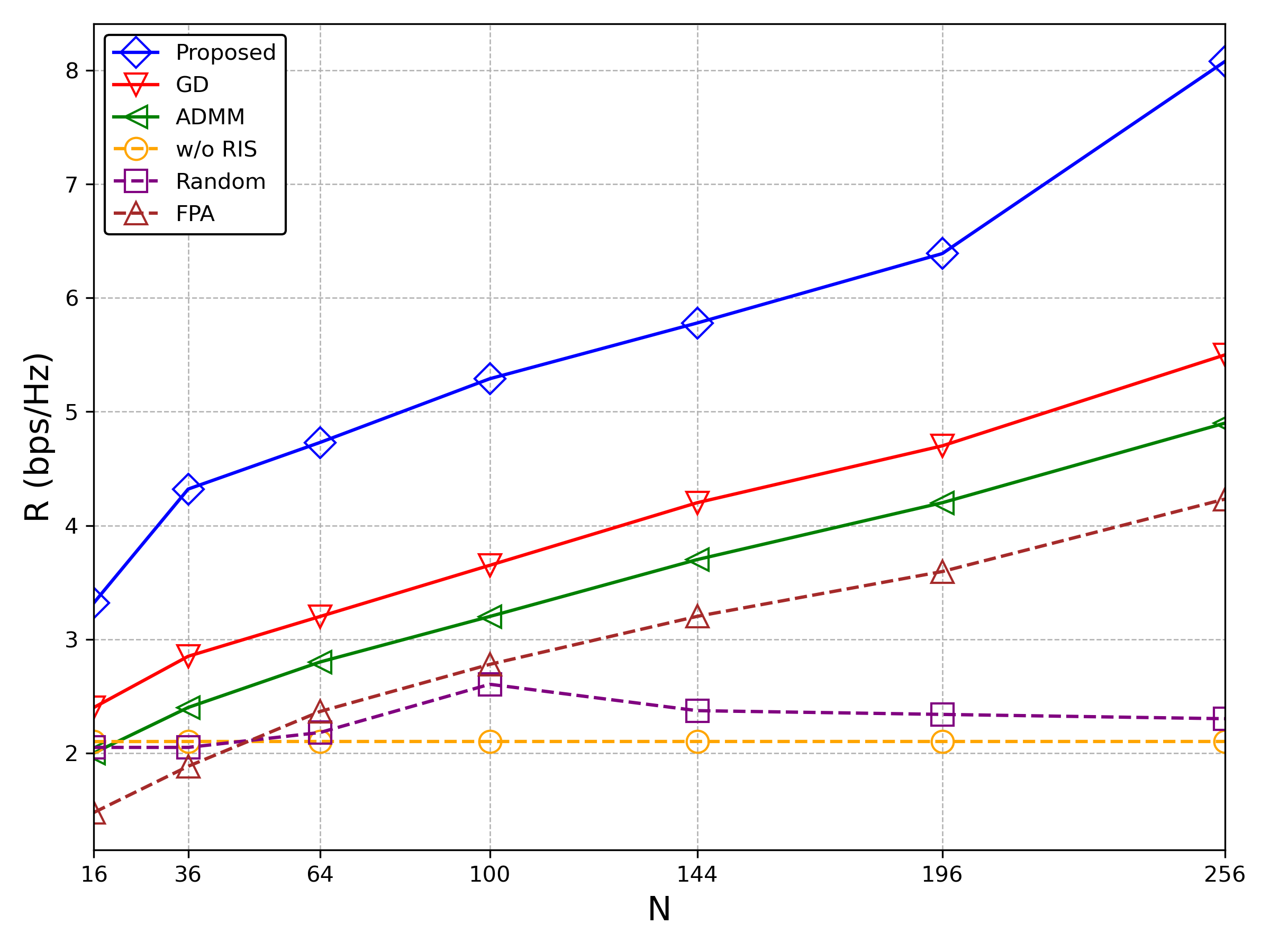}
    \caption{The number of RIS reflecting elements versus minimum achievable rate $R$.}
    \label{fig:RISN}
\end{figure}

Fig.~\ref{fig:RISN} shows the achievable rate versus the number of RIS elements $N$. depicts the variation of the minimum achievable rate with respect to the number of RIS elements N, ranging from 16 to 256.
A notable observation is that the proposed field-driven alternating optimization (FAO) exhibits a steep acceleration in achievable rate for $N > 64$. This sharp rise is because a larger RIS aperture enhances interference suppression by enabling finer control over reflected sidelobes, and simultaneously amplifies coherent signal aggregation, since the reflected wavefront can be more precisely aligned toward the desired user.
In contrast, the w/o RIS baseline remains nearly flat because the system lacks any reflective enhancement—the received power depends solely on the direct path, and increasing N provides no benefit.
The FPA curve starts below the w/o RIS case, but later surpasses it as 
N grows. This is because, without RIS assistance, the FPA cannot avoid interference nor locate the strongest radiation zones; however, when the RIS aperture becomes sufficiently large, its enhanced reflection gain compensates for the lack of antenna mobility.
\emph{A larger RIS not only strengthens the effective signal power through coherent combining but also enables fine-grained control of interference fields. Meanwhile, the FA remains indispensable. Its spatial adaptability complements RIS phase reconfiguration, jointly achieving robust interference suppression and efficient energy focusing in large-aperture deployments.}

\begin{figure}[t]
    \centering
    \includegraphics[width=0.8\linewidth]{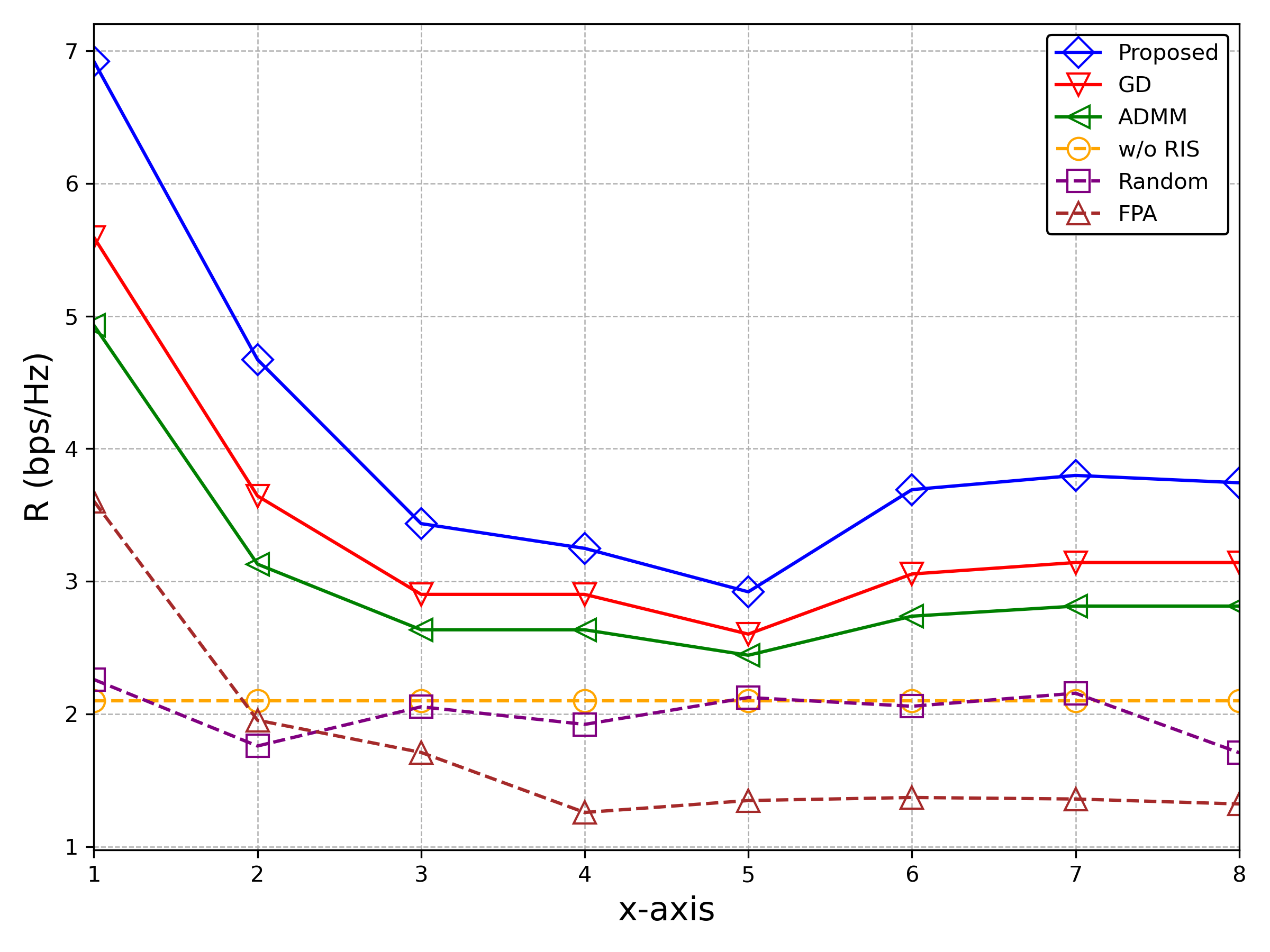}
    \caption{RIS x-axis versus Achievable rate versus minimum achievable rate $R$.}
    \label{fig:RISX}
\end{figure}

Fig.~\ref{fig:RISX} illustrates the minimum achievable rate as a function of the RIS horizontal position along the $x$-axis, while the receiver is fixed at the origin and the transmit power is kept constant. The RIS is gradually moved away from the user toward the BS, covering a range of $x\in[1,8]m$. The achievable rate first decreases and then increases as the RIS moves along the $x$-axis. This phenomenon arises from the multiplicative path-loss effect between the user–RIS and RIS–BS links. FAO consistently outperforms the baselines across all positions. Its advantage lies in 3DGRF modeling, which captures spatial field variations caused by RIS relocation and dynamically re-optimizes both the RIS phase and the FAS position to maintain coherent energy focusing and interference suppression. In contrast, the w/o RIS case remains nearly constant because it lacks reflective enhancement. The Random baseline exhibits noticeable fluctuations because its discrete random phase configuration leads to inconsistent constructive and destructive interference, underscoring the importance of precise RIS phase control. The FPA baseline remains below the proposed method because fixed antennas cannot spatially adapt to field changes.
\emph{Placing the RIS near either endpoint enhances both the coherent reflection gain and the controllability of interference.}

\begin{figure}[t]
    \centering
    \includegraphics[width=0.8\linewidth]{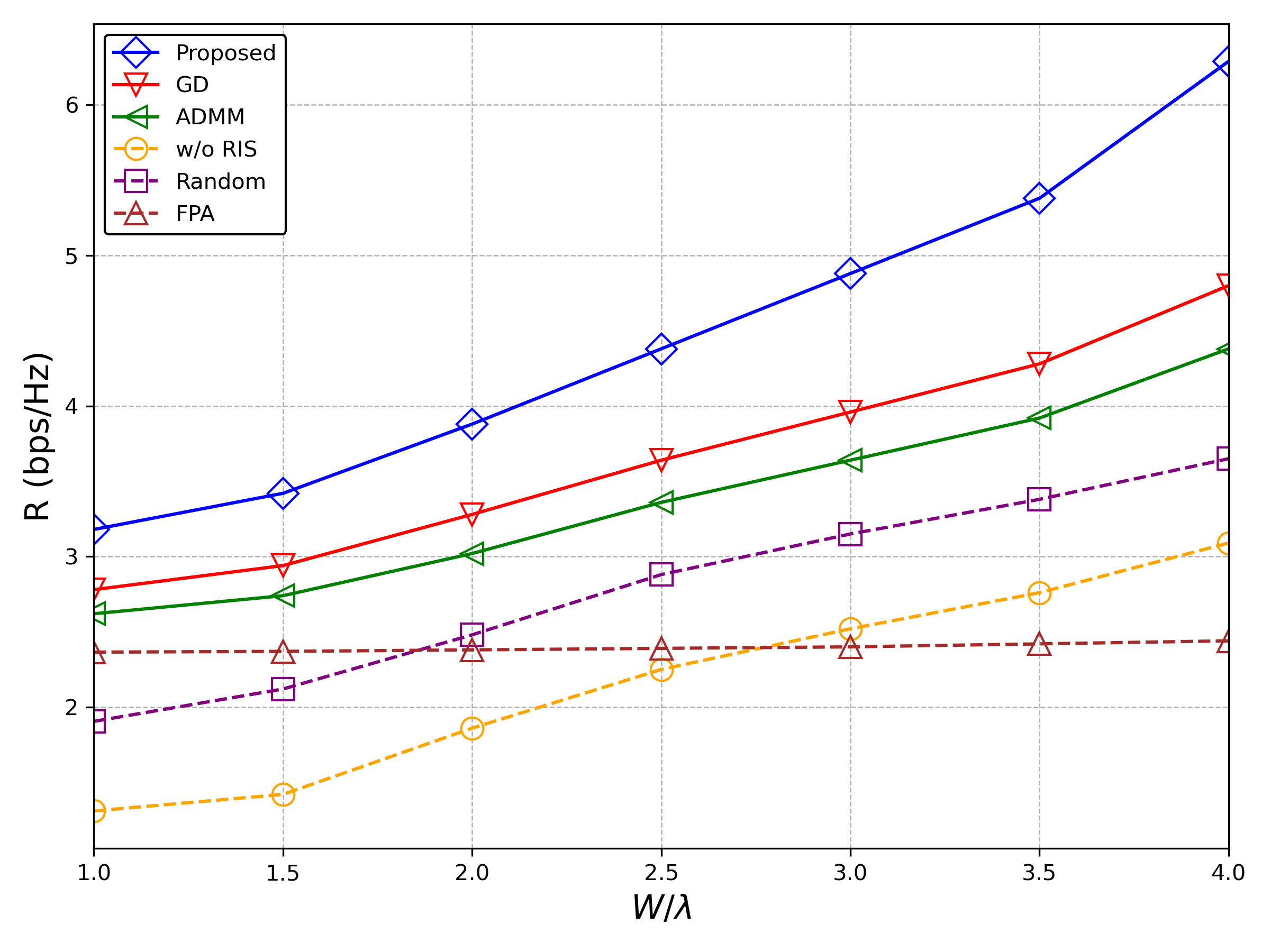}
    \caption{The normalized FAs’ movable range $W/\lambda$ versus minimum achievable rate $R$.}
    \label{fig:FASA}
\end{figure}

Fig.~\ref{fig:FASA} illustrates the achievable rate versus the normalized FAS movement range $W/\lambda$. As $W/\lambda$ increases, all methods exhibit steady performance improvement, indicating that a larger FAS movement region provides more spatial degrees of freedom. With a wider exploration space, the antenna can select positions with stronger radiation intensity and reduced interference, thereby improving the received SINR. The proposed FAO achieves the highest growth rate and the steepest slope across all ranges. Its 3DGRF model accurately captures the fine-scale variation in the electromagnetic field, enabling the antenna to move adaptively to the most favorable positions. By contrast, GD and ADMM exhibit slower improvement because they rely on discrete CSI estimation and iterative convex approximations. As the movement region enlarges, its optimization space grows exponentially. The w/o RIS baseline improves only marginally since FAS mobility alone cannot fully compensate for the absence of reflective gain.
\emph{Expanding the FAS movable range increases the system’s spatial adaptability, enabling joint exploitation of field peaks and interference nulls. However, as the search space grows, optimization efficiency becomes the performance bottleneck. Therefore, the combination of advanced field-driven algorithms and flexible FAS hardware is essential for achieving high-rate, interference-resilient communication in next-generation 6G networks.}

\section{Conclusions}\label{s6}
This paper has presented a novel field-driven optimization framework for the FAS-RIS system based on 3DGRF modeling. By explicitly representing the electromagnetic energy distribution using differentiable Gaussian primitives, the proposed method transforms the conventional channel-driven paradigm into a radiation-field-based optimization process. This unified formulation enables joint optimization of FAS positions and RIS phase shifts with low pilot overhead and computational complexity, achieving real-time adaptability in fast-fading and spatially non-stationary environments.
Comprehensive simulations validated that the proposed framework achieves superior spectral efficiency, convergence speed, and latency performance compared with traditional channel estimation–based or statistical modeling approaches. The results confirm that 3DGRF modeling not only enhances the physical interpretability of the optimization process but also bridges the gap between wireless field reconstruction and communication configuration, offering a scalable and energy-efficient solution for next-generation 6G systems.

\appendices

\section{}
\label{appendix:ssim}

Let $\mathbf{\xi}_m$ denote the spatial observation position of the $m$-th receiver antenna. The ground-truth electromagnetic field $E_{\mathrm{gt}}(\mathbf{\xi}_m)$ is obtained from the theoretical model as
\begin{equation}
E_{\mathrm{gt}}(\mathbf{\xi}_m) =
\sqrt{P} \, h_{km} +
\sum_{n=1}^{N} \sqrt{P} \, h_{kn} e^{j\theta_n} h_{nm},
\label{eq:gt_field_appendix}
\end{equation}
where $h_{km}$, $h_{kn}$, and $h_{nm}$ denote the direct and RIS-reflected channel coefficients.

The SRN aims to reconstruct $\tilde{E}(\mathbf{\xi})$ from the sampled field, and is optimized by minimizing the hybrid loss function
\begin{equation}
\mathcal{L} = (1 - \eta)\|E_{\mathrm{gt}}(\mathbf{\xi}) - \tilde{E}(\mathbf{\xi})\|_2^2
+ \eta \big(1 - \mathrm{SSIM}(E_{\mathrm{gt}}(\mathbf{\xi}), \tilde{E}(\mathbf{\xi}))\big),
\label{eq:srn_loss_appendix}
\end{equation}
where $\eta \in [0,1]$ is the balancing coefficient. 
The SSIM is defined as
\begin{equation}
\mathrm{SSIM}(x, y) =
\frac{(2\mu_x \mu_y + C_1)(2\sigma_{xy} + C_2)}
{(\mu_x^2 + \mu_y^2 + C_1)(\sigma_x^2 + \sigma_y^2 + C_2)},
\label{eq:ssim_def_appendix}
\end{equation}
with $\mu_x$, $\mu_y$, $\sigma_x^2$, $\sigma_y^2$, and $\sigma_{xy}$ denoting mean, variance, and covariance, respectively.

\section{}
\label{appendix:gaussian}

A continuous spatial field $E(\mathbf{s})$ can be approximated by $M$ Gaussian kernels:
\begin{equation}
E(\mathbf{\xi}) \approx \sum_{i=1}^{M} G_i(\mathbf{\xi}).
\label{eq:gaussian_sum_appendix}
\end{equation}

For an isotropic emitter centered at $\boldsymbol{\mu}_i$ with variance $\sigma_i^2$, we have
\begin{equation}
|G_i(\mathbf{\xi})| = A_i
\exp\!\left(-\frac{\|\mathbf{\xi}-\boldsymbol{\mu}_i\|^2}{2\sigma_i^2}\right).
\label{eq:isotropic_gaussian_appendix}
\end{equation}
For anisotropic spatial distributions, the covariance matrix $\boldsymbol{\Sigma}_i$ replaces $\sigma_i^2$, yielding
\begin{equation}
|G_i(\mathbf{\xi})| = A_i
\exp\!\left(-\frac{1}{2}(\mathbf{\xi}-\boldsymbol{q}_i)^{\mathrm{T}}
\boldsymbol{\Sigma}_i^{-1}(\mathbf{\xi}-\boldsymbol{q}_i)\right).
\label{eq:anisotropic_gaussian_appendix}
\end{equation}

Extending to complex-valued representation, each Gaussian primitive is defined as
\begin{equation}
G_i(\mathbf{\xi}) = A_i e^{j\zeta_i}
\exp\!\left(-\frac{1}{2}(\mathbf{\xi}-\boldsymbol{q}_i)^{\mathrm{T}}
\boldsymbol{\Sigma}_i^{-1}(\mathbf{\xi}-\boldsymbol{q}_i)\right),
\label{eq:complex_gaussian_appendix}
\end{equation}
where $A_i$ and $\psi_i$ denote the amplitude and phase, respectively.

Hence, the Gaussian primitive in~\eqref{eq:complex_gaussian_appendix} corresponds to the analytical form used in the main text.

\bibliographystyle{IEEEtran}
\bibliography{IEEEabrv,reference}

\end{document}